%% file: main.tex
\documentclass[manuscript,screen,acmlarge,authorversion]{acmart}

\AtBeginDocument{%
  }

\setcopyright{acmcopyright}
\copyrightyear{2018}
\acmYear{2018}
\acmDOI{XXXXXXX.XXXXXXX}

\acmConference[Conference acronym 'XX]{Make sure to enter the correct
  conference title from your rights confirmation emai}{June 03--05,
  2018}{Woodstock, NY}
\acmPrice{15.00}
\acmISBN{978-1-4503-XXXX-X/18/06}

\usepackage{soul}
\usepackage{subcaption}
\usepackage{xcolor,colortbl}

\newcommand{\imagepath}[1]{cropped-figures/#1}
\newcommand{\rr}[1]{\textcolor{black}{#1}}

\usepackage{soul}
\usepackage{xcolor}

\definecolor{highlighter}{HTML}{fff100}
\sethlcolor{highlighter}

\begin{document}

\title{The Prompt Artists} 


\author{Minsuk Chang}
\affiliation{%
  \institution{Google, Inc.}
  \city{Seattle, Washington}
  \country{United States}}
\email{misukchang@google.com}

\author{Stefania Druga}
\affiliation{%
  \institution{Google, Inc.}
  \city{Mountain View, California}
  \country{United States}}
\email{druga@google.com}

\author{Alex Fiannaca}
\affiliation{%
  \institution{Google, Inc.}
  \city{Seattle, Washington}
  \country{United States}}
\email{afiannaca@google.com}

\author{Pedro Vergani}
\affiliation{%
  \institution{Google, Inc.}
  \city{London}
  \country{United Kingdom}}
\email{pvergani@google.com}

\author{Chinmay Kulkarni}
\affiliation{%
  \institution{Google, Inc.}
  \city{Atlanta, Georgia}
  \country{United States}}
\email{ckulkarni@google.com}

\author{Carrie Cai}
\affiliation{%
  \institution{Google, Inc.}
  \city{Mountain View, California}
  \country{United States}}
\email{druga@google.com}

\author{Michael Terry}
\affiliation{%
  \institution{Google, Inc.}
  \city{Seattle, Washington}
  \country{United States}}
\email{michaelterry@google.com}

\renewcommand{\shortauthors}{Chang et al.}

\input{sections/abstract.tex}


\begin{teaserfigure}
\includegraphics[width=\textwidth]{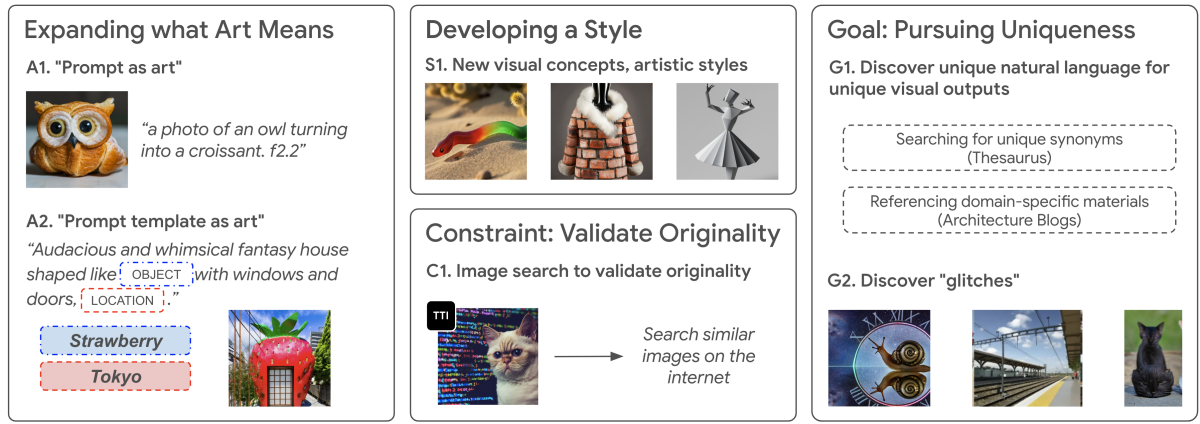}
\caption{Prompt artists develop descriptive text-based prompts that are rendered by text-to-image models. Highly skilled prompt artists will develop 1) distinct visual concepts and styles (S1), 2) prompts that can also serve as titles of the art piece (``prompts as art'', A1), and 3) ``prompt templates'' (A2), which encapsulate specific visual concepts to be customized by others. Artists strive to discover unique \textit{natural language} that produces unique \textit{visual outputs} (G1), and/or model ``glitches'' (G2) that can be elevated to artistic styles in their own right. Finally, some prompt artists validate the novelty of their work by conducting an image search for similar images (C1).}
\label{fig:teaser}
\end{teaserfigure}

\maketitle

\input{sections/new_intro}
\input{sections/relatedwork}
\input{sections/interview}
\input{sections/artists_new}

\input{sections/themes}
\input{sections/discussion}
\input{sections/conclusion}

\begin{acks}
\end{acks}

\bibliographystyle{ACM-Reference-Format}
\bibliography{references}

\end{document}

%% file: sections/abstract.tex
\begin{abstract}
This paper examines the art practices, artwork, and motivations of prolific users of the latest generation of text-to-image models. Through interviews, observations, and a user survey, we present a sampling of the artistic styles and describe the developed community of practice around generative AI. We find that: 1) the text prompt \textit{and} the resulting image can be considered collectively as an art piece (\textit{prompts as art}),
and 2) \textit{prompt templates} (prompts with ``slots'' for others to fill in with their own words) are developed to create \textit{generative art styles}. We discover that the value placed by this community on unique outputs leads to artists seeking specialized vocabulary to produce distinctive art pieces (e.g., by reading architectural blogs to find phrases to describe images). We also find that some artists use ``glitches'' in the model that can be turned into artistic styles of their own right. From these findings, we outline specific implications for design regarding future prompting and image editing options.

\end{abstract}

%% file: sections/new_intro.tex
\section{Introduction}
Advances in text-to-image(TTI) models have led to significant improvements in the quality of computer-generated, synthetic images \cite{NEURIPS2021_49ad23d1, ho2022cascaded}. A new generation of text-to-image models enable the creation of high-fidelity images via descriptive text prompts by leveraging advances in large language models ~\cite{DALLE279:online,Midjourn12:online,saharia2022Imagen,yu2022Parti,Rombach_2022_CVPR}. 
With broadening access to these models, communities of practice have emerged, enabling people to share designs, prompts, and example images. For instance, there are now tools to help people write prompts~\footnote{https://promptomania.com/prompt-builder/}, and even marketplaces for successful prompts~\footnote{https://promptbase.com/shop}.

Prior work has examined the phenomenon of computer-generated art in a variety of contexts  
~\cite{10.1145/3326338,aguera2017ArtInAI,mazzone2019ArtCreativityAI}. For example, as an art historian analyzing the AI-assisted art movement, Mazzone et al.~\cite{mazzone2019ArtCreativityAI} describe how the artist's role has adapted to include pre-curation, tweaking, and post-curation. 
More recently,~\citeauthor{hertzmann2020ComputersArt} argues that text-to-image (TTI) models like DALL·E do not themselves create art, but that the artists and technologists who apply them as tools are the ones creating art \cite{hertzmann2020ComputersArt}. With the emergence of this new class of models, which are capable of producing extremely high quality images from textual descriptions (e.g., \cite{PartiPat10:online, ImagenTe79:online, DALLE279:online}), 
we are motivated to understand how this new technology is being adopted and used by creators.

In this research, we provide a snapshot of a vibrant community of \rr{art} practice that has arisen around text-to-image models, \rr{ sharing insights into the ingenuity and creativity of the users of these models.\footnote{We thank our anonymous reviewers for the specific phrase recognizing our contribution.}}. Within a US-based technology company that has produced its own TTI models, we sent a survey to the TTI models users to gain a basic understanding of how and why they are used. We also interviewed and observed 11 prominent users of these models \rr{who are using the models as an art medium}, recruiting from survey respondents and by directly asking prominent users. \rr{They have each generated thousands of images with both the internal and other publicly available models, and are actively sharing their creations in multiple communities.} In studying \rr{the artistically-driven} members of this local community, we sought to understand their practices, their artifacts, and their motivations for engaging extensively with the models. For the purposes of this research, we scope our inquiry to studying interfaces that only accept text as input (recognizing that a wide variety of model capabilities and interfaces are available, including those that enable more fine-grained editing of images). We restrict our scope to text-only interfaces because these were the first interfaces available for models such as DALL-E, and these text-based interfaces have seen considerable use by the public and our internal users. 

Our study reveals that users of these models have developed a range of artistic styles, including origami figurines, fashion (e.g., dresses) made out of materials like bricks, and reality ``mash-ups'' that create hybrids of animals or of fruits and vegetables (see Figures \ref{fig:dresses}, \ref{fig:origami}, \ref{fig:mashups}, \ref{fig:hybrid-animals}). However, we also found that the artistic outputs of this community of users are not limited to the images themselves. For example, the \textit{prompt itself} is an important output, and a piece of the art: a parsimonious, descriptive prompt accompanying the image is seen as a virtuous goal beyond just the image, as it simultaneously acts as a ``title'' for, and description of, the art piece (see Figure \ref{fig:teaser}, A1). Similarly, a \textit{prompt template}---a text prompt with one or more empty ``slots'' for others to fill in---is considered an art piece all on its own (see Figure \ref{fig:teaser}, A2). Among other characteristics, a well-designed prompt template has the property of encapsulating an artistic vision that can nonetheless be customized by future users of that prompt template.

Our results also reveal the lengths some users go to when searching for unique, distinctive outputs. In particular, some creators turn to thesauri or online, domain-specific blogs (e.g., architectural blogs) in search of vocabulary that elevates the model output beyond the ordinary. This focus on vocabulary suggests that capable TTI model artists may also benefit from being highly skilled with natural language. Another creator  explicitly seeks unique model outputs, but through identification of ``glitches'' that can be elevated to styles all on their own. For example, this latter artist found the model did not  render reflections in mirrors perfectly, and explored this concept through a number of pieces (see Figure \ref{fig:teaser}, G2).

Finally, we find that the artists interviewed place a premium on originality, with some turning to image search to validate that their outputs are, in fact, unique.

In sum, this paper presents results from a survey and interview study of heavy users of TTI models, making the following contributions:
\begin{itemize}
    \item \textbf{Usage summary}: From survey data from 161 respondents, we find that when they use a model, 20\% of respondents report using a TTI model for one or more hours at a time, indicating fairly sustained use of these models by a sizable portion of the community surveyed.
    \item \textbf{Sample styles}: We provide a sampling of artistic styles developed by study participants to contextualize the types of outputs being produced by new text-to-image models.
    \item \textbf{Prompt as art}: We find that the prompt itself is often considered a part of the artistic output (in addition to the actual image), with artists pursuing a goal of creating parsimonious, descriptive input prompts.
    \item \textbf{Prompt templates as art}: We discover that artists also produce \textit{prompt templates} to encapsulate a unique visual concept that others can customize.
    \item \textbf{Natural language mastery for visual language artistry}: We describe how TTI artists seek unique natural language in an attempt to elevate their pieces beyond the norm.
    \item \textbf{Glitches as art}: We show how some artists look for ``glitches'' that can be reliably transformed into new styles.
    \item \textbf{Validating originality}: We describe artists' concerns in validating their outputs as original, and how they currently validate through image search.
\end{itemize}

Together, these findings suggest new directions for interactive interfaces and aids for prompt-centric uses of TTI models: 1) Methods and tools to help users locate novel language and capabilities of the model, 2) aiding users in validating the originality of their outputs, and 3) reifying the notion of a prompt template into a standalone computational artifact that supports richer interaction by users of the template. \rr{Importantly, while our results derive from a study of internal TTI models, the implications for design are generally applicable to use of any TTI model (e.g., the notion of a prompt template is useful for any TTI model, as it captures a particular artistic vision in a portable, yet customizable. form).}

In the rest of the paper, we review related work, describe our study method, present results from the survey and interview study, and conclude with a discussion that draws implications for design from the study data.

%% file: sections/relatedwork.tex
\section{Related Work}

Advances in deep learning have led to the development of generative machine learning models capable of producing both images that are highly realistic and images that are highly creative in both existing and novel artistic styles \cite{aiart2022Cetinic}. For example, Generative Adversarial Networks (GANs) \cite{gan2014Goodfellow} learn to generate images and simultaneously distinguish real and fake images to generate highly realistic images. Many variations of the GAN architecture have been investigated (e.g., \cite{Zhu_2017_ICCV, Karras_2019_CVPR, gan2018Brock}). A particularly relevant variation of this architecture is the Creative Adversarial Network (CAN) from Elgammal et al. \cite{aican2017Elgammal}, designed to generate images with novel artistic styles. This artwork was subsequently featured in multiple exhibitions \cite{AICAN73:online} where human observers could not distinguish the CAN-generated art from human-authored artwork. Outside of GANs,  Gatys et al. \cite{gatys2015NeuralStyle} introduced a method to apply learned artistic styles to random images (a technique now known as Neural Style Transfer \cite{jing2020NeuralStyle}). Additionally, work originally designed to make convolutional neural networks more explainable, now referred to as Deep Dreams, became popular for generating art \cite{GoogleDeepDream81:online} due to its ability to generate psychedelic versions of images \cite{mordvintsev2015DeepDreams}. While these techniques allowed for generating images with creative and novel artistic styles \cite{aican2017Elgammal}, none provided significant affordances to end-users for controlling what was generated outside of the training data scope.

Mansimov et al. \cite{mansimov2015ImgFromCaption} addressed this issue by showing that a generative model could produce novel images from natural text when conditioned on image captions. As text-to-image models rely on language modeling techniques, recent advances in the scaling of large language models \cite{devlin2018Bert, raffel2020exploring} have enabled the development of correspondingly large text-to-image models with impressive results. The most recent of these models include: DALL·E \cite{pmlr-v139-ramesh21a, DALLECr11:online} and DALL·E 2 \cite{ramesh2022DALLE2, DALLE279:online}, Stable Diffusion \cite{Rombach_2022_CVPR, StableDi88:online}, Midjourney\cite{Midjourn12:online}, Parti \cite{yu2022Parti, PartiPat10:online}, and Imagen \cite{saharia2022Imagen, ImagenTe79:online}. 

Fueled by the latest advances in text-to-image models, current image generation applications are becoming mainstream. With this broader adoption comes the question of how these new models' capabilities impact art practices, which we examine in this paper.


\subsection{AI in Creativity Support Tools \& Human-AI Co-Creation}

AI tools have played a prominent role in creativity support tool (CST) research \cite{frih2019CSTinHCI, hwang2022Creative}. AI-based CST systems have been produced to support artistic generation in domains such as fashion and product design \cite{Sbai_2018_ECCV_Workshops, jeon2021Fashion, quanz2020machine}, music creation \cite{mccormack2019Music, louie2020Music, huang2020AISong}, drawing \cite{davis2015Drawing, davis2016Drawing,  oh2018Drawing, karimi2019Drawing}, visual design and story-boarding \cite{zhao2020IconGen, shi2020Storyboarding}, and storytelling \cite{hodhod2016Storytelling, perone2019Chatbot}.
Most of these tools support the artistic implementation process as either production aids (i.e., tools that perform most of the work of generating the art, e.g., generative text-to-image models) or as execution aids (e.g., AI-powered brush tools in drawing applications) \cite{chung2021Intersection}. 

Hwang et al. \cite{hwang2022Creative} further characterize how these tools apply AI models in the creative process as falling into four general categories: \textit{editors} (facilitate execution of processes), \textit{transformers} (aid in changing existing content), \textit{blenders} (combine 2 or more content sources), and \textit{generators} (produce novel content). In this framing, the large-scale TTI models that this work is focused on fall into the \textit{generators} category.

Additionally, research in the related field of Human-AI Co-Creation (a sub-field of Mixed-Initiative Co-Creativity \cite{yannakakis2014mixed}) is highly relevant. In the \textit{Library of Mixed-Initiative Creative Interfaces} \cite{spotoLibrary}, Spoto et al. proposed a framework to understand mixed-initiative co-creation as a process involving seven potential actions: \textit{ideate}, \textit{constrain}, \textit{produce}, \textit{suggest}, \textit{select}, \textit{assess}, and \textit{adapt}. 
Muller et al. \cite{muller2020mixed} extended this framework to generative AI applications, while Grabe et al. further simplified this extension and characterized four primary interaction patterns concerning GAN applications: curating, exploring, evolving, and conditioning \cite{grabe2022towards}. In our work, we observe similar themes, especially the notion of users of TTI models feeling like they are using the models to explore, and acting as curators of their outputs.

Work in this field has also identified core challenges in creating human-AI co-creation systems. For example, Chung et al. \cite{chung2021gestural} identified the limited ability to control the output of generative AI models and proposed using gestural input to constrain/guide the model output. Likewise, Buschek et al. \cite{buschek2021nine} identified a set of nine challenges system designers could encounter when developing human-AI co-creative systems. In the context of TTI models, they identified challenges of \textit{invisible AI boundaries} (``A (generative) AI component imposes unknown restrictions on creativity and exploration'') and \textit{conflicts of territory} (``AI overwrites what the user has manually created/edited'') as particularly salient. Participants in our study encountered similar challenges of identifying what the models are capable (and not capable) of, and in building upon prior inputs to the system.


Building on this prior CST research, we examine the emergent practices and goals that have evolved in concert with the latest generation of TTI models.

\subsection{Generative AI as an Artistic Medium}

Given the rapid advancement of AI models for generating images with ever more creative and novel styles, art historians and technologists have been actively discussing how to conceptualize AI-assisted art in relation to other artforms. Experts in these communities have disagreed as to whether generative models should be considered artists in-and-of themselves \cite{mazzone2019ArtCreativityAI} (as with the famous sale of \textit{Portrait of Edmond Belamy} auctioned by Christie's in 2018 \cite{Thefirst77:online}) or whether they should be considered as merely a tool employed by artists. Hertzmann \cite{hertzmann2018ComputersArt, hertzmann2020ComputersArt} argues that generative AI models are similar to the camera as it relates to the art of photography: a tool the enables the art. Hertzmann further theorizes that ``art is an interaction between social agents,'' and generative AI models are therefore best considered an agent in this interaction. Agüera y Arcas \cite{aguera2017ArtInAI} provides a similar argument with an in-depth discussion of the similarity between the emerging field of AI-generated art and photography, particularly surrounding the historical reaction of painters to the introduction of the camera. Grba \cite{grba2021brittle, grba2022DeepElse} provides a framework to critically evaluate art created with a generative AI model. In this framework, he echoes arguments above, and critiques what he refers to as ``the ever-receding artist'': the repeated occurrence of technologists referring to models as artists, thereby minimizing the contribution of the human who employed the model to create art. Finally, Browne \cite{brown2022AIArtist} explored what it means to be an ``AI artist'' and proposed the framing of an AI artist as a \textit{bricoleur} (building upon \cite{grba2019forensics}), saying, ``\textit{Bricolage} is common to generative art, where ideas are developed through playful experimentation with existing tools and techniques.''


In our paper, we present an analysis of interviews with artists who employ a large TTI model for the generation of their art, highlighting the work of three of these artists and provide context around their motivations and goals. In our results, we find thematic alignment with perspectives advanced by Hertmann, Agüera y Arcas, and Grba above: While the models can produce surprisingly high quality output, dedicated users of these models employ the models as tools to explore specific themes and concepts. Accordingly, they have intentionally developed processes (e.g., locating domain-specific terminology) to improve their ability to achieve their individual goals.

\subsection{Diffusion and Auto-Regressive Models}
\rr{Diffusion models (e.g., Imagen~\cite{ImagenTe79:online} or DALL-E~\cite{DALLE279:online}) are trained by gradually adding noise to an image, until all of the image is completely noise. The model then learns to reverse the noising process to generate the original image. In this way, a diffusion model learns to synthesize an image from noisy images, and is capable of generating images from arbitrary ``noise.'' For text-to-image models, the models are also trained (conditioned) on text inputs~\cite{nichol2021glide}, allowing the model to produce an image from a noisy image input and text input, where the resulting image bears a resemblance to the input text.}

\rr{Autoregressive models, such as Parti~\cite{PartiPat10:online}, treat text-to-image generation as a sequence-to-sequence problem, akin to machine translation or other language modeling tasks. In the case of TTI models, the ``translation'' is from text to image (i.e., text tokens to image tokens).}

\rr{In our study, participants used both types of models.}

%% file: sections/interview.tex
\section{Study Design}
To understand current practices, motivations, and goals when using modern text-to-image models, we sent a survey to internal users of two internal TTI models to collect basic information about their use of these models (e.g., time spent using the models, motivations, desired capabilities, and prompting strategies). We also interviewed and observed 11 \rr{power} users of TTI models (8 identifying as Male and 3 identifying as Female) in a 50-minute study to uncover their motivations and practices. The latter participants were prolific users of one or more TTI models. Models used by the participants in the study are anonymized for review, but are of the same basic capability as state of the art text to image generation models such as Imagen \cite{ImagenTe79:online}, Parti \cite{PartiPat10:online}, and DALL-E 2 \cite{DALLE279:online} which are described in more detail in our related work.



\subsection{Participants}
For the survey, participants were recruited from an internal chat channel dedicated to TTI models (where the internal chat channel has thousands of members) and internal TTI model mailing lists (with hundreds of members). From the survey respondents, we identified interview candidates who had reported having created artwork over more than ten sessions and having spent more than five hours in the previous week using a TTI model. To create a pool of participants, we recruited eight of these latter respondents, and further recruited three prominent artists in the internal artist community to participate. These three artists are quite visible in the internal artist community, and have shared their unique artwork collections within that community. \rr{Participants were also actively engaged in external communities, sharing knowledge, expertise, and artwork.} Participants were given a 60 USD gift card for participation.

\subsection{Interview structure}
The study consisted of four parts intended to understand participants' practices. Each participant was first asked to create an image of their choice to allow the researcher to observe their natural practices.
In the second part of the study, participants were asked to reflect on 1) an artwork they were proud of, 2) a piece they found most successful, and 3) the piece that was least successful. 
In the third part of the interview, participants were asked to reflect on someone else's work by examining only the prompt, and specifically asked to either improve or change the prompt in their style. 
In the last part of the interview, participants were asked to discuss envisioned uses for these text-to-image models.

Interviews were conducted remotely. We recorded the shared screen and automatically generated transcripts for the interviews. We constrained our focus to interfaces that only use text prompts as input to the models. In addition, some participants voluntarily shared their collection of generated artwork after the interview.

\subsection{Qualitative Data Analysis}

For the qualitative analyses, the authors analyzed the video transcriptions and also noted comments on participants non-verbal interactions. The final corpus of automatically generated transcripts was 164 pages (60614 words). The first two authors each reviewed the transcripts data independently, looking for ways of explaining the artistic practices~\cite{miles1984drawing}. In this process, the authors separately analyzed each transcript to extracted salient themes, and independently generated hypotheses and points of discussions~\cite{braun2006using}. Using these data, all authors participated in two rounds of interpretation sessions to arrive at the primary themes reported in this paper, \rr{and resolve any discrepancies and disagreements. During the interpretation sessions, authors also analyzed the prompts and the images created by the study participants to identify unique artistic styles and practices. These sessions were inspired by existing analysis practices from qualitative media analysis \cite{altheide2012qualitative}}. 

%% file: sections/artists_new.tex
\section{Survey Results}

We received 161 responses to the survey. Of these responses, 160 answered the question, ``At present, why are you using the [TTI] models?'' Of the responses to this question, 79 (49\%) indicated they use the models to create art, 33 (21\%) reported using it as part of their creative work pipeline, and 126 (79\%) indicated their use was curiosity-driven (not work-related).

In the survey, we also asked participants to estimate the length of time they work with a TTI model when they use one (``When you interact with a model, how much time do you typically spend interacting with it?''). We received 157 responses to this question, with 20\% of respondents indicating that they use a model for one or more hours at a time when they use it (11\% reporting using it for 1-2 hours at a time, 9\% using it for 2 or more hours at a time), 53\% indicating they use a model for 10 minutes to an hour, and 27\% reporting use for less than 10 minutes.


When asked about observed strengths and weaknesses of the various models they've interacted with, survey responses indicated a number of desired capabilities, such as the ability to render text in images, the ability to have more control over spatial arrangements, and the ability for models to handle complex prompts. We also asked participants for desired capabilities when interacting with the model. Seventy six percent of the respondents requested a ``how to build a prompt'' guide, and 75\% desired the ability to fork and remix images, especially for spatial refinement. Seventy six percent of the responses also indicated they would like features like bookmarking, and the ability to directly share outputs to  internal chat groups or social media. As we will see in the artist spotlights below, there is a clear social component to working with these TTI models for our study participants.

Finally, 63\% of the responses desired greater control over the model, such as the ability to assign specific values to each of the prompt words. 


When asked to provide prompting techniques they have learned, 
common themes for the strategies included 1) producing specific art styles and eras, such as ``impressionist style'', 2) use of keywords that describe camera lenses and aperture (e.g., ``DSLR photo'',``3D render'',``24 mm, f8, ISO1000''), and 3) domain-specific terms (e.g., ``Line Art'', ``black and white'').

\section{Prompt Artists: Styles, Motivations, Practices}

In this section, we provide a sampling of the vibrant internal TTI artistic community by spotlighting the work of three highly active creators; Shai Noy, Irina Blok, Dan Smith.  \footnote{In the text, we use the terms ``artists,'' ``creators,'' and ``study participants'' interchangeably. We denote the three spotlighted artists as A1, A2, and A3, and other participants by a participant number (e.g., P1, P2, ..., P8).}. For these three artists, we describe and present examples of the styles they have developed and summarize their artistic motivations and goals. We then provide a summary of motivations, styles, and practices observed across the 11 interview participants. In the section that follows, we describe salient high-level, emergent themes arising from the interviews and observations. We want to credit the artists whose artwork are featured, and they had expressed the desire to be associated with their artwork. We use their full name in places where the artwork appears.

\subsection{Shai Noy: The Explorer (A1)}
\begin{figure}[t]
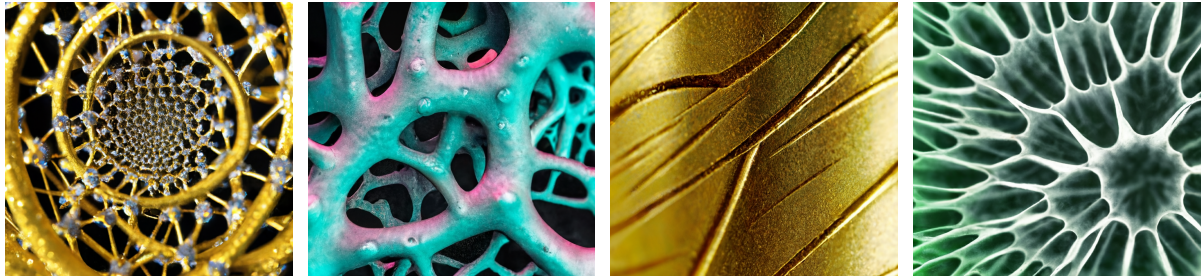
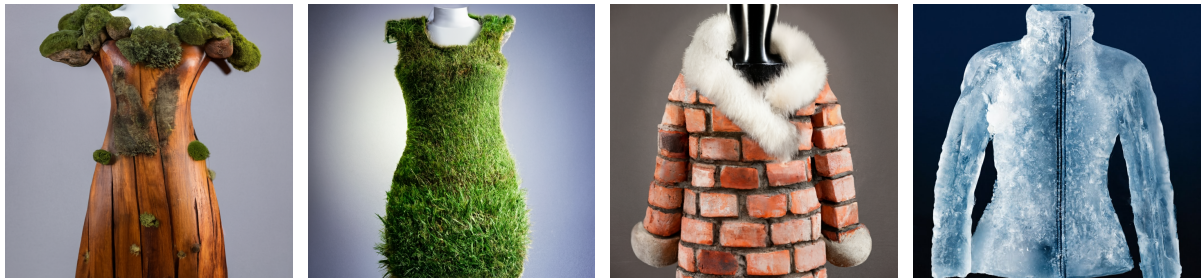


    \begin{subfigure}[b]{\textwidth}
        \includegraphics[width=0.24\linewidth]{\imagepath{A1_-_macro_micro_photo_1.png}}
        \hspace{\fill} 
        \includegraphics[width=0.24\linewidth]{\imagepath{A1_-_macro_micro_photo_4.png}}
        \hspace{\fill} 
        \includegraphics[width=0.24\linewidth]{\imagepath{A1_-_macro_micro_photo_2.png}}
        \hspace{\fill} 
        \includegraphics[width=0.24\linewidth]{\imagepath{A1_-_macro_micro_photo_3.png}}
        \caption{Super macro photography} 
        \label{fig:macro}
    \end{subfigure}

    \vspace*{0.2cm} 

    \begin{subfigure}[b]{\textwidth}
        \includegraphics[width=0.24\linewidth]{\imagepath{A1_-_A_beautiful_dress_carved_out_of_dead_wood_with_lichen_and_mushrooms,_on_a_mannequin._High_quality,_high_resolution,_studio_lighting.png}}
        \hspace{\fill} 
        \includegraphics[width=0.24\linewidth]{\imagepath{A1_-_A_beautiful_dress_made_out_of_grass_and_dirt,_on_a_mannequin._High_quality,_High_resolution,_Studio_lighting.png}}
        \hspace{\fill} 
        \includegraphics[width=0.24\linewidth]{\imagepath{A1_-_A_beautiful_winter_coat_made_out_of_a_brick_fireplace,_on_a_mannequin._High_quality,_high_resolution,_studio_lighting.png}}
        \hspace{\fill} 
        \includegraphics[width=0.24\linewidth]{\imagepath{A1_-_A_beautiful_winter_jacket_made_out_of_ice,_on_a_mannequin._Studio_lighting,_high_quality,_high_resolution.jpg}}
        \caption{Dresses in various unusual materials}
        \label{fig:dresses}
    \end{subfigure}
    \caption{Selections from Shai Noy (A1 - The Explorer)} 
    \label{fig:artist1}
\end{figure}

Shai Noy is a software engineer with no training in the visual arts or design. However, they have produced thousands of images with TTI models. The styles developed by this artist include ``super macro photography'' images (i.e., extremely close-up views of objects, Figure \ref{fig:macro}), and fashion (dresses, suits) made out of unusual materials, such as wood, grass, brick, or ice (Figure \ref{fig:dresses}).

Elements of both discovery and community were emphasized as rewarding for this artist, such as being the first to explore particular concepts and the ability to share discoveries: 
``Everything is more fun when you can share it'' and, ``Art doesn't live in a vacuum, nobody starts from scratch, everything is based on something else. I am proud of being able to recognize the potential'' (A1).

\subsection{Irina Blok: The Art Director (A2)}
\begin{figure}[t]
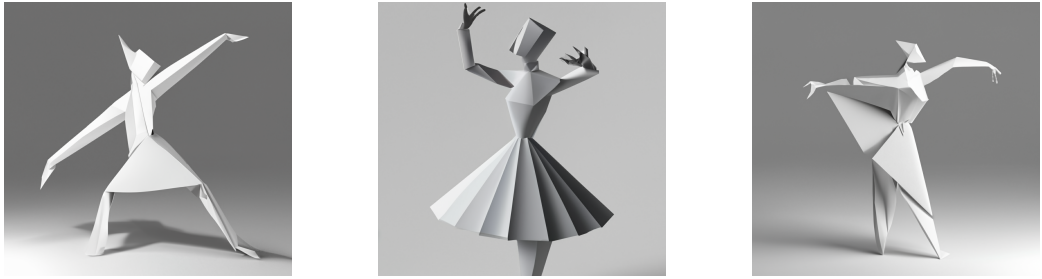
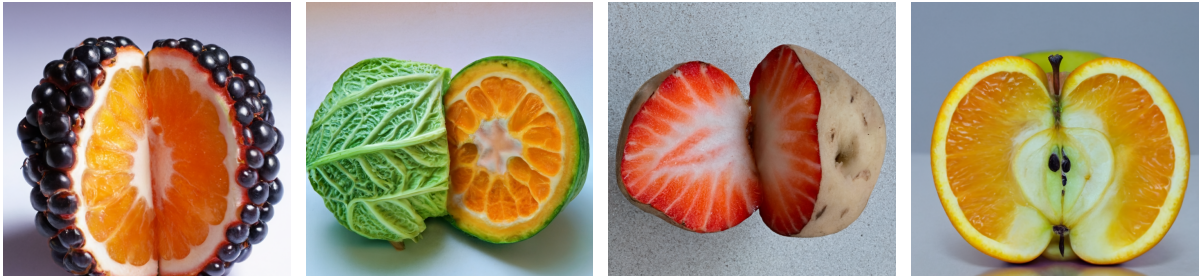

    \begin{subfigure}[b]{\textwidth}
        \hspace{\fill} 
        \includegraphics[width=0.24\linewidth]{\imagepath{A2_-_origami_dancer_1.png}}
        \hspace{\fill} 
        \includegraphics[width=0.24\linewidth]{\imagepath{A2_-_origami_dancer_2.png}}
        \hspace{\fill} 
        \includegraphics[width=0.24\linewidth]{\imagepath{A2_-_origami_dancer_3.png}}
        \hspace{\fill} 
        \caption{Origami dancers} 
        \label{fig:origami}
    \end{subfigure}
    
    \vspace*{0.2cm} 
    
    \begin{subfigure}[b]{\textwidth}
        \includegraphics[width=0.24\linewidth]{\imagepath{A2_-_Two_objects_1.png}}
        \hspace{\fill} 
        \includegraphics[width=0.24\linewidth]{\imagepath{A2_-_Two_objects_2.png}}
        \hspace{\fill} 
        \includegraphics[width=0.24\linewidth]{\imagepath{A2_-_Two_objects_3.png}}
        \hspace{\fill} 
        \includegraphics[width=0.24\linewidth]{\imagepath{A2_-_Two_objects_4.png}}
        \caption{Reality mash-ups} 
        \label{fig:mashups}
    \end{subfigure}
    \caption{Selections from Irina Blok, (A2 - The Art Director)} 
    \label{fig:artist2}
\end{figure}

Irina Blok is a designer who has done visual work for many years using stock images and applications like Photoshop. This artist's styles include origami dancers (Figure \ref{fig:origami}) and reality ``mash-ups,'' such as sliced produce that contains different textures internally (e.g., a sliced head of cabbage that reveals a cross-section of an orange, Figure \ref{fig:mashups}).

Driving these styles is a passion for developing ``impossible objects, something we haven’t seen before, mashups'' (A2). They also seek to create images that differ significantly from the images it was trained on: ``The further away the generated images are from what it was trained on, the more the satisfaction'' and, ``they're [the resulting images] also defying the rules of its training set, like defying [...] gravity. [..] A creative prompt breaks that'' (A2).

Importantly, this artist's output also includes \textit{prompt templates} that describe a particular, parameterized image or visual concept, such as this prompt for generating a house: \textit{Audacious and whimsical fantasy house shaped like <object> with windows and doors, <location>}. These prompt templates are carefully crafted to reliably produce pleasing results for others. We describe this concept more fully in a later section.


Notably, when working with the text-to-image models, A2 considers the model to be akin to an artist itself, with themselves the art director. In this relationship, A2 acknowledges a certain lack of control: ``[I] don’t have full control, and there’s beauty in this'' (A2). At the same time, they also consider the model to be a tool: ``[The model] is a brush [...] you just learn to speak its language'' (A2). With these dual views (model as an artist, model as a tool), they note that ``the hardest part is [the] conceptual aspect, being skillful with [the] prompt'' and that ``it’s a thought exercise, it’s not a visual exercise'' (A2). It’s really about how to make people think like an artist''  (A2). In this latter sentiment, this participant was specifically speaking to the need to think like an artist in formulating a prompt, as opposed to formulating a prompt as if one was talking to a machine.



\subsection{Dan Smith: The Social Commentator (A3)}

\begin{figure}[t]
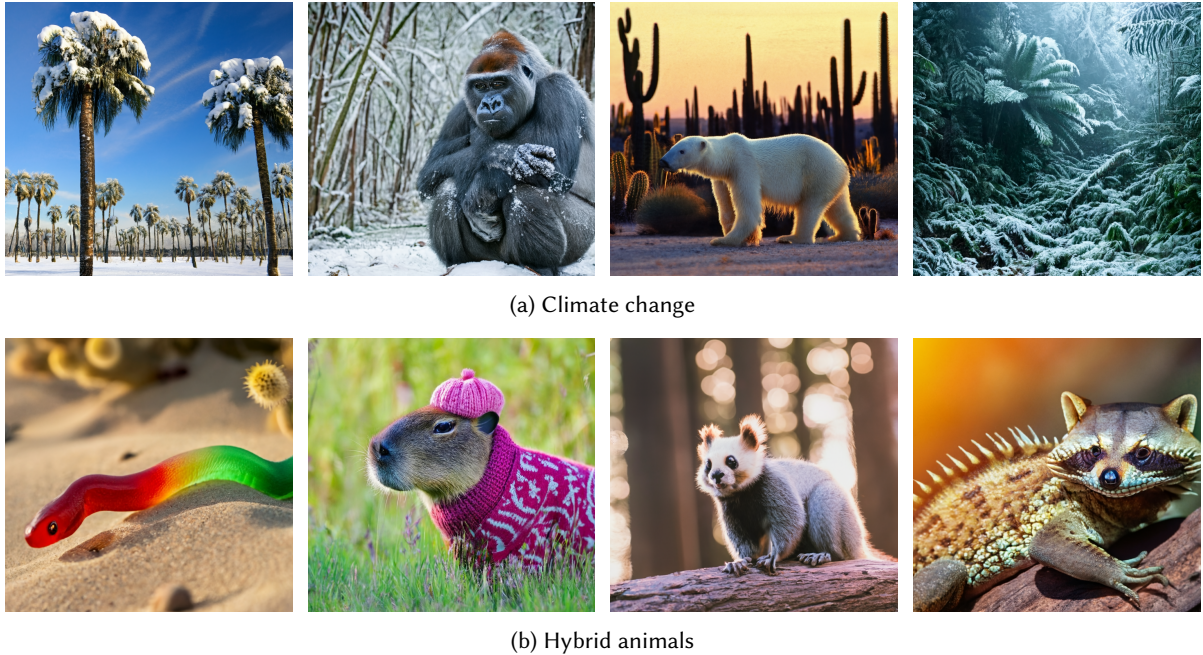

    \begin{subfigure}[b]{\textwidth}
        \includegraphics[width=0.24\linewidth]{\imagepath{A3_-_Climate_1.png}}
        \hspace{\fill} 
        \includegraphics[width=0.24\linewidth]{\imagepath{A3_-_Climate_2.png}}
        \hspace{\fill} 
        \includegraphics[width=0.24\linewidth]{\imagepath{A3_-_Climate_3.png}}
        \hspace{\fill} 
        \includegraphics[width=0.24\linewidth]{\imagepath{A3_-_Climate_4.png}}
        \caption{Climate change} 
        \label{fig:climate-change}
    \end{subfigure}
    
    \vspace*{0.2cm} 
    
    \begin{subfigure}[b]{\textwidth}
        \includegraphics[width=0.24\linewidth]{\imagepath{A3_-_Hybrid_1.png}}
        \hspace{\fill} 
        \includegraphics[width=0.24\linewidth]{\imagepath{A3_-_Hybrid_2.png}}
        \hspace{\fill} 
        \includegraphics[width=0.24\linewidth]{\imagepath{A3_-_Hybrid_3.png}}
        \hspace{\fill} 
        \includegraphics[width=0.24\linewidth]{\imagepath{A3_-_Hybrid_4.png}}
        \caption{Hybrid animals} 
        \label{fig:hybrid-animals}
    \end{subfigure}
    \caption{Selections from Dan Smith (A3 - The Social Commentator)} 
    \label{fig:artist3}
\end{figure}

Dan Smith comes from a background in visual media, and is partially driven by the desire to deliver a message about the climate crisis, in order to facilitate change and awareness: ``[I'm] not just having fun ... but [actually] making something that has some power'' (A3). In working towards this goal, they want to make ``something that you look at, and it makes you feel something'' while also making ``something that people would want to look at'' (A3). A3's image styles align with these overall goals: They have a number of pieces that put nature in ``situations where you wouldn't see it'' (Figure \ref{fig:climate-change}) and have created numerous ``hybrid animals'' (Figure \ref{fig:hybrid-animals}). This particular style---hybrid animals---is also inline with a motivation to create images that ``would be hard to [...] visualize or create, if you were [...] a really skilled Photoshop artist'' (see Figure \ref{fig:artist3}).

A3 also heavily considers the quality of the image when assessing its outputs: it must have near expert-level composition, photorealism, and/or artistry. When describing the artwork they created and how they created them, they noted, ``I think the ones that I pick out are [...] standouts for various reasons, and just [...] like what I said, [...] composition photorealism, artistic quality'' (A3). Consistent with their emphasis on overall image quality, they have found ways to address undesirable outputs. For example, in generating images that include animals, they found they needed to adopt a specific strategy to create aesthetically pleasing images: ``I would do `Tall Grass' a lot because early on I discovered that limbs and fingers and paws can get a little wonky'' (A3). The ``tall grass'' addition was their creative strategy to hide feet or paws and suggests an understanding of model limitations, but also a sense of how to cope with these limitations.

\subsection{Summarizing Motivations, Styles, and Practices}

One of the primary motivations for interacting with the models was that participants found them fun---the models' output quality enabled people to feel creative, and they were generally interested in interacting with this new class of model. Some people also noted that the model enabled them to engage in their domain interests in a new way. For example, one participant said that the models allowed them to explore their interest in Swiss trains with the model, while another found it compelling to try to create new forms of currency (e.g., new types of coins). One participant used it to create the equivalent of clip art for presentations: ``I use [TTI] model when I need some [...] clipart to use in my presentation, and [TTI] model would be amazing for that...'' (P2).

In comparing the artists spotlighted above to the other participants, one notable difference in practices is that the spotlighted artists placed a particular emphasis on exploring specific themes in depth (e.g., origami figurines). The other participants did not pursue concepts with the same rigor or depth.

%% file: sections/themes.tex
\section{Themes: Art Beyond Images, Discovering Unique Points, and Validating Originality}

Across our interviews and observations, a number of themes emerged. First, the notion of what constituted the final artistic output was not always just an image: Some creators consider the \textit{prompt itself} as part of the art. Similarly, a \textit{prompt template} can be considered an artistic output. Second, there was a clear desire for creators to discover new styles possible with the models. This focus on originality extended to one creator going so far as to conduct an image search on successful outputs to validate their originality. We unpack these themes further in this section: 1) Prompt as art, 2) Prompt template as art, 3) Discovering new capabilities of the model, and 4) Validating originality.

\subsection{Prompt as Art}

\rr{For some participants, the prompt itself was part of the overall art, and thus worthy of attention. For these participants, it was important to ``[create] aesthetically pleasing images'' and ``[develop] art concepts'' that \textit{were inherently tied to prompts}.
We detail these motivations and behaviors below.}

While generating aesthetically pleasing, ``glitch-free'' images was a common goal of the creators, other goals were also present in their practices. For some participants, the prompt itself was part of the overall art, and thus worthy of attention 1) on its own, and 2) as it relates to the image: ``It's part of the aesthetic'' (P3), where the prompt is ``like a title of the piece, but you don't get to choose it independently'' (P3). Hinted at in this last comment is the notion that while a prompt could also serve as a title of the art piece, there is a clear dependency on, and perfunctory role for, that prompt as well: The prompt serves as the source material for the model generating the resultant image. Given this dependency, finding a prompt that produces the desired result \textit{and} that can serve as a title for the piece can be challenging, but rewarding when it happens.

\begin{figure}[t]
    \captionsetup[subfigure]{justification=centering}
    \begin{subfigure}[t]{0.4\textwidth}
    \centering
        \includegraphics[width=0.6\linewidth]{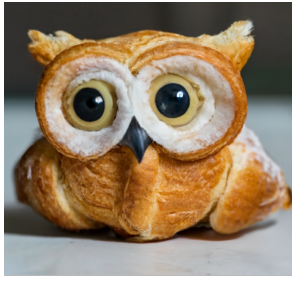}
        \subcaption{``a photo of an owl turning into a croissant. f2.2''} 
        \label{fig:owlcroissant}
    \end{subfigure}
    \begin{subfigure}[t]{0.4\textwidth}
        \centering
        \includegraphics[width=0.6\linewidth]{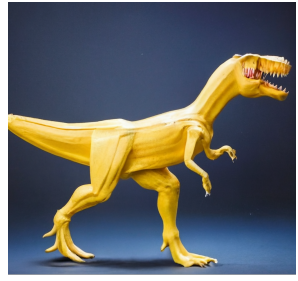}
        \subcaption{``a photo of a bananasaurus rex. a bananasaurus rex has the legs and arms of a banana and the body and head of a tyrannosaurus rex. sigma 85mm f2.2. studio lighting.''} 
        \label{fig:bananasaurus}
    \end{subfigure}
    \caption{Artist Ian Fischer (P3) considers (a) as a successful art work given the brevity of the prompt, while (b) a failed attempt because the prompt is too long to get the image they want}
    \label{fig:prompt-as-art}
\end{figure}

This same artist (P3) further described the prompt's role as 
``communicating the image, and the idea of the image, and how I got it all at the same time'' (P3). 
This quote sheds light on how art with TTI models can, in some sense, be considered multimodal (text and image) for both the artist and viewer: The prompt and the final output combine into a single, mutually-reinforcing, art piece.
Seen in this light, one can consider the \textit{prompt as art} itself---a well-crafted prompt that creates a compelling image but also accompanies that image, saying something about the image. See Figure \ref{fig:prompt-as-art}a for an example of ``prompt as art,'' as well a prompt that wasn't able to achieve this same level of aesthetic (Figure \ref{fig:prompt-as-art}b).

Diving deeper into this theme, we 
observed that the artists believe the ``concept'' is a critical component of an artwork. In an internal blog article, A2 describes, ``There’s a common misconception art is largely about drawing and painting skills. Art is not only about how something looks, it’s about what it says, it tells a story, and has a concept. Art can surprise, provoke, teach, delight and inspire. Art is not just about drawing, art is a way of thinking.'' A1 also suggested ``the unit of shareable artwork is not necessarily a specific image but maybe it's the whole exploration of the concept of those images.''


In pursuing ``prompt as art,'' we observed participants impose different constraints when creating their prompts, with some wanting it as descriptive as possible, while others attempting to make it as simple as possible. One participant even discovered delight in their accidental discovery that their ``random string'' produced beautiful output, and named that prompt as their proudest achievement. In this latter case, the pride comes from the joint pairing of the random string and the beautiful output---without the context of the random string, the image has less value, as the random string reveals an unexpected feature of the model.



\subsection{Prompt Templates As Art}

\begin{figure}[t]
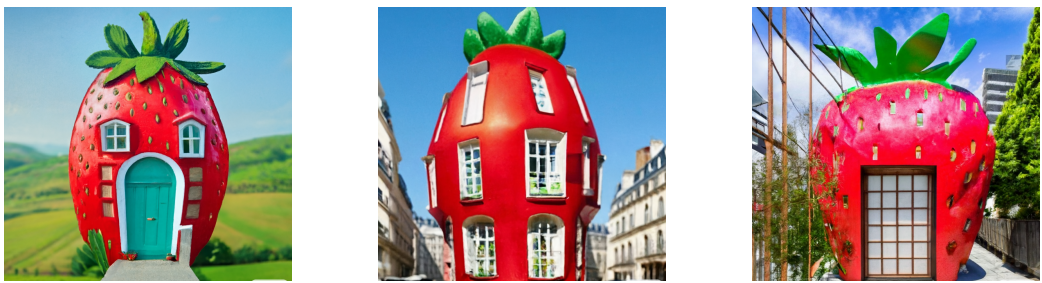

        \hspace{\fill} 
        \includegraphics[width=0.24\linewidth]{\imagepath{strawberry_countryside.png}}
        \hspace{\fill} 
        \includegraphics[width=0.24\linewidth]{\imagepath{strawberry_paris.png}}
        \hspace{\fill} 
        \includegraphics[width=0.24\linewidth]{\imagepath{strawberry_tokyo.png}}
        \hspace{\fill} 
        \caption{Example of the prompt ``Audacious and whimsical fantasy house shaped like <object> with windows and doors, <location>''. All images above use ``strawberry'' as the shape, with the location varied (``countryside'', ``Paris'', ``Tokyo'').} 
        \label{fig:house_template}
\end{figure}

In a spirit similar to ``prompt as art,'' five artists sought to produce \textit{prompt templates}. A prompt template is an image description with ``slots'' for someone else to fill in. For example, we previously noted this prompt template created by A2: \textit{Audacious and whimsical fantasy house shaped like <object> with windows and doors, <location>} (see Figure \ref{fig:house_template} for example outputs of this prompt template).

\rr{These prompt templates leverage the capabilities of the models as well as the lightweight, accessible features of prompts. More specifically, when an artist identifies a compelling composition, they can create a text prompt that allows others to create a similar composition, but with their own unique customization to it. 
The ease with which the templates can be shared also introduces social motivations for producing and distributing templates (e.g., to participate and contribute to a larger community of practice). We elaborate on these points below. }

Prompt templates have a number of key features:

\begin{itemize}
    \item They are tightly coupled to and represent a particular artistic concept, vision, or composition (such as a ``whimsical fantasy house''). These features make prompt templates conceptually richer than words or phrases used to specify stylistic characteristics of the image (e.g., ``35mm'' or ``watercolors'', which may be used to produce a particular effect, but don't specify a larger composition).
    \item Others can make use of prompt templates by filling in the blanks. The richness of the models means that the templates guide the overall generated image, but users' unique input can yield a diverse variety of outputs.
    \item There is the intent for the prompt template to provide consistently high quality and delightful output when used by others.
\end{itemize}

Unpacking these concepts in the context of the example prompt above, the skeleton of this prompt embodies a particular (visual) concept: A fantasy home in a given location. Someone making use of this prompt can customize it through two key variables: A shape for the house and a location for the house. While these are seemingly simple variables to customize, the template gives the user great flexibility in terms of the final outputs produced by the model: Any number of shapes can be provided in any number of locations (in fact, the user is free to substitute any text in the slots they wish).

Simultaneously, the prompt cues the model to the \textit{types} of output to produce, as well as details to guide the generation. The phrase ``audacious and whimsical fantasy house'' defines desired attributes of the house, while the specification of ``with windows and doors'' provides additional details that should be included in the generated house. These details in the prompt help increase the reliability of the model output and reduce the likelihood that the downstream user of the prompt template needs to experiment further with the overall prompt structure and content to obtain a good output. 

A noteworthy characteristic of these templates and the text-to-image models is the rich interplay that results between the original template and the phrases the user substitutes in the template. For example, a particular choice of location can profoundly influence the resulting house generated by the model---the model does not simply generate the same type of house and situate it in a different location. Instead, the choice of location can directly interact with the other parts of the prompt. For example, creating a strawberry house in the ``countryside'', ``Paris'', or ``Tokyo'' yields qualitatively different outputs for the house, with the house style meshing more naturally in the chosen location (e.g., having styles of windows more typical for a European building for the house in Paris, and doors more typical of Japan for Tokyo---see Figure \ref{fig:house_template}). These interactions between the template and the users' choices enable a diverse variety of outputs that allow a user to explore a wide range of ideas.

To produce prompt templates, one participant described a process whereby they would 1) input a prompt, 2) identify high-quality outcomes, and 3) rewrite the prompt to try to produce that same outcome again. This process could involve several iterations until they get a reliable prompt. Once a prompt is working, participants would sometimes remove content to make it more succinct. For example, they would first emphasize a specific characteristic (like ``high resolution'') by applying many descriptors representing a similar effect (such as ``high res'', ``DSLR'', ``crystal clear'', ``photo realistic''), then start removing the repeated qualifiers in the prompt until the prompt could reliably generate similar outputs to the original, verbose version. 

As with the notion of a ``prompt as art,'' the creation of these prompt templates can also be considered an artistic outcome in and of itself: The prompt author must first develop a compelling concept, then ensure it has enough capacity to enable people to produce their own unique creations within the frame of the concept. When done well, a prompt template has the quality of attracting users (because of the compelling output produced by the prompt) \textit{and} continually delighting users with the outputs produced using their unique input.

\rr{Elaborating further on this notion of ``prompt template as art,'' a prompt template represents a particular \textit{artistic vision}, with the prompt text capturing the overall design, composition, and aesthetic intentions of the author. Notably, because the artistic vision is captured in natural language text, a prompt template can be used \textit{across} TTI models: Users can customize the template as desired, then tweak the prompt to produce the desired output using whatever model they have at their disposal.}

\subsection{Discovering Unique Points in the Model's Latent Landscape}
A common goal of many artists was to discover new capabilities of the model that others had not yet found. As one participant put it, they tried to ``break'' the model, pushing it to its limits. As with prompt templates, these newly discovered capabilities are often shared so others can apply the concept in their own images. For example, an artist may identify the ability of the model to produce physical sculptures out of unusual materials or to create origami-like figurines (as A2 did). Once a concept and ability have been identified, others can build on the core idea and create their own images or permutations. \rr{In particular, participants felt proud when they ``made the model do X (new concept)'', especially when they discovered a model capability for the first time in the community in which they heavily engage, as the discovery could influence the discourse in the community. 
We expand on these points below.}

new words to steer the model in new directions. For example, one artist mentioned referencing architectural blogs to learn that domain's vocabulary so it could be applied to their prompts (A2). Another artist mentioned turning to a thesaurus to enrich their vocabulary for the prompt. As one concrete example, A1 (the ``Explorer'') likes to employ ``unusual words'' to steer the model, such as ``intricate'' instead of ``detailed'', explaining: ``If you choose common words, then you get a bit of an uninspiring result quite often. But if you use something a bit more unusual, then you really narrow down [...] the set [of images], and you're going to get into things that use this less common word'' (A1).

What's noteworthy here is that in the pursuit of novel imagery, creators would carefully research and choose \textit{words} to produce specific \textit{visual} outcomes: \textit{What you say and how you say it is critical to producing high-quality imagery with text-to-image models}, requiring TTI users to \textit{enrich their natural language vocabulary} in order to develop and skillfully execute a unique \textit{visual language}.

Driving these practices of discovering new capabilities was a clear desire to push the model away from ``average'' outputs and get it into more unique spaces. In this sense, the artists are navigating and charting the vast latent space of the model and sharing back the most interesting places discovered. When a new, unique output space was discovered, artists expressed a certain satisfaction in having discovered that space: `` A good rule of thumb is to be more descriptive than not [...] if you see that something is lacking, then you try to add more descriptors that will encourage the model, in this direction [...] sometimes you just have to reword a sentence or move something from one sentence to another [...] it's mostly like binary search'' (A1); `` I think your choice of vocabulary is very interesting because you want a large tree. But then, instead of saying a large tree you say mature tree [...] Where do these [...] choices come from? They just come from trial and error'' (A2).

\begin{figure}[t]
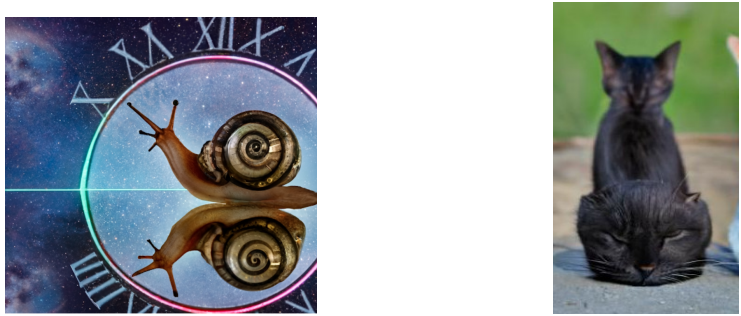

        \hspace{\fill} 
        \includegraphics[width=0.26\linewidth]{\imagepath{P1.png}}
        \hspace{\fill} 
        \includegraphics[height=0.26\linewidth]{\imagepath{single_cat_back.jpg}}
        \hspace{\fill} 
        \caption{Artwork by Paul Emmerich(P8). Examples of ``breaking the model'': the artist seeks out and elevates ``glitches'' in the model's output to create new styles. The artist takes advantage of the model's imperfect reflections (on the left) while making creative use of its tendency to transform the backs of cats into cat faces (see the bottom part of the cat's back, which looks like a cat's face).} 
        \label{fig:breaking_the_model}
\end{figure}

In seeking unique spaces, one participant (P8) expressly hunted for interesting imperfections; instead of \textit{avoiding} glitches, they \textit{sought} glitches. For example, this participant found one of their prompts created imperfections in its output for ``a hybrid of a clock and a snail on an infinite mirror. Steampunk. DSLR photo. astrophotography'' (Figure \ref{fig:breaking_the_model}). Here, A4 explores the imperfections of the infinite mirror: it's ``technically a failure but still amazing'' (P8). P8 also discovered that the model sometimes has trouble generating the backs of things (like cats) and developed that behavior into its own art style (Figure \ref{fig:breaking_the_model}).

\subsection{Validating Originality}

One question that often arises on the topic of human-in-the-loop, AI-generated art is one of creativity and originality---how much can be attributed to the AI versus the person~\cite{hertzmann2020ComputersArt}. Our participants also struggled with this question, with some seeking to ensure their outputs were novel. \rr{Participants wanted to validate the originality of the artistic concept, but could only check the originality of the artifact.} For example, one participant described a practice of using Google's image search on compelling outputs to ensure originality: After producing a creative output, they use image search to search for that image (or similar images) to ensure it is, in fact, original. This participant mentioned they do this in part because they are sensitized to the fact that machine learning models can sometimes memorize portions of the training data. Given this, they want to ensure their output is \textit{not} in the training data.

%% file: sections/discussion.tex
\section{Discussion}
In this research, we have provided a snapshot of the emerging artistic scene enabled by the latest generation of text-to-image models. Our interviews and observations document the types of imagery artists are developing with these new models, as well as an enhanced understanding of the types of results creators seek beyond the images themselves (e.g., prompts and prompt templates as important artifacts in their own right).

In this section, we discuss some implications deriving from artists' goals of seeking novelty, validating originality, and producing reusable prompt templates.  In reviewing design implications for these models, we note that there are countless ways the input and editing interfaces could be improved for interacting with these models (for example, one can find many proposals and working demos online). In this discussion, we restrict ourselves to interfaces that only take text as input, with no post-processing of the image (such techniques that allow in-filling of regions by the model). We limit our discussion to this input modality in part because a number of the artists interviewed were seemingly attracted to the simplicity (and challenge) that this single input modality provides.

\subsection{Aiding Novelty}
In our interviews, the use of thesauri and domain-specific blogs (e.g., architectural blogs) illustrate the desire of artists to identify unique terms that help them produce results that rise above the average. Embedding these search capabilities directly into the tooling could be useful (e.g., quick access to a thesaurus or an embedded search engine). Pushing this idea further, there may also be opportunities to make use of large language models (LLMs) and/or the TTI model's training data to help surface salient terms for a given topic. For example, an LLM may be able to generate terms-of-art for architecture using a prompt like this: \textit{Here are some terms specific to describing the architectural design of a house: 1)}. In our own test with an LLM \footnote{Specific model anonymized for review}, this prompt yielded terms like \textit{entryway}, \textit{foyer}, \textit{entry hall}, and \textit{elevation}. To use the training data itself to identify new spaces, it may be possible to collect terms associated with a particular topic, such as ``house,'' then identify other terms that are more frequently found near that topic in the training data compared to the rest of the training data (e.g., using a method such as term frequency-inverse document frequency (TF-IDF)).

To help users understand how unique their word choices are, one could also visualize the input prompt with respect to how common each individual input token is. For templates, one could also show what the common terms would be for filling in the blanks to help people move beyond those common terms to find more distinct, unique inputs. One possible outcome in helping people find more novel inputs is that their inputs may lie outside of the training distribution, leading to unpredictable results. Providing feedback through mechanisms like visualizations (e.g., showing frequency in the training data) could help users better understand these types of issues should they arise. 

\subsection{Validating Originality}
As mentioned, there is a desire to validate the originality of the image produced by the model. Streamlining the process of doing an image search with an online search engine is one obvious way to address this issue. However, as is the case in aiding novelty (described above), there may be opportunities to take advantage of the training data itself. More specifically, in addition to external search, one could also search for the closest images in the training data to the image produced.

While the above mechanisms would help validate the outputs generated, there may also be opportunities to help validate the originality of the \textit{input}. For example, one might be able to search the training data to identify which parts of a prompt exist within the training data.

\subsection{Materializing Prompt Templates}
Prompt templates provide a way for an artist to derive a new style, then share it with others so they can produce their own unique images. One could imagine embracing this practice and transforming a prompt template into its own first-class interface.

For example, one could imagine allowing users to provide multiple inputs for each slot of a prompt template, then generating the cross-product of all the inputs. One could also make use of the fact that the prompt text for the templates is tokenized into vectors. Specifically, by supplying two different inputs for the same slot, embedding vectors for each input could be obtained and then automatically interpolated between the embeddings to produce a spectrum of outputs. For example, for the shape of a house (in the previous house prompt template), the user could provide two inputs: strawberry and apple. The system could then produce the embedding vectors for those inputs, then interpolate between those embeddings to create a series of images that morph from a strawberry to an apple. However, one thing to keep in mind with these interpolations is they are occurring in the text space rather than image space---the model will be interpolating not between shapes (per se) but between the \textit{linguistic concepts} of strawberries and apples (which may still produce an interesting morphing between these conceptual entities).

\subsection{Prompts and TTI Models as an Art Medium}
One the primary ways the three spotlighted artists (A1, A2, and A3) distinguished themselves from others was their perception of TTI models as an art medium, with a clear focus on exploring the capabilities and limits of medium itself, rather than only on individual outcomes. In this spirit, they embraced 
the limitation of not being able to edit the images directly, accepting the text-only input as a defining feature and characteristic of the medium. For example, A2 expressed ``there's beauty in it'' when describing the prompting interaction with the models. In contrast, other participants focused more on the outcome, and expressed desire for additional features, such as direct editing of the generated images. 
Embracing the TTI models as-is further reinforces the idea that text prompts are part of the artwork, rather than simply a means to an end. 

This observation suggests that the design implications for TTI models can be considered from multiple perspectives: prompt-only artists, and creative professionals using the models to achieve specific goals.
Creative professionals with a specific design goal may require and request 
specific features that offer more fine-grained control, perhaps with tools to help understand model behavior. For example, features related to directly editing the generated images were among the most frequently requested features in 
the survey. 
\rr{Given that non-design experts can also learn to use TTI models to quickly demonstrate and visualize artifacts, we hypothesize that this may facilitate more active and iterative communication between design professionals and their clients, with clients more directly engaged in the creative process (as opposed to the more traditional pipeline-like design workflow). If this proves to be true, feedback and collaboration features could become more important}.

\subsection{Limitations}
Our participant pool was drawn among employees of one large US-based corporation, and does not cover other possible ways that culture, community, and collaboration might shape use of TTI models (e.g. on social media). Also, since our analysis was episodic rather than longitudinal, we do not document how artistic prompting strategies may evolve within individuals. For the interactions we \textit{could} observe, observing participants' interactions with the TTI model does not definitively indicate their conceptions of how the model works or how best to prompt it. We also acknowledge that different models behave in different ways due to their structure, training data, and the design of the interface. \rr{In that regard, we also acknowledge some of the findings might be model dependent, and are specific to the models used in the study. Similarities and differences in art forms and art practices might be observable with different models and communities. Another limitation is that while our participants are actively involved in multiple communities, 
we did not ask participants about their experiences with other models or other communities in depth.}

%% file: sections/conclusion.tex
\section{Conclusion}
In this paper, we have described a unique moment in time: Recent text-to-image models have given rise to an exceptionally vibrant community of practice, complete with new ideas about what constitute notable outcomes (e.g., new styles, prompts as art, and prompt templates as art). As the larger artistic and creator community adopts these new models and forms of art, there are clear ways in which the tools can improve to better support desired practices including: 1) helping to discover and create novel outputs, 2) providing methods to validate novelty, and 3) elevating the notion of a prompt template into a standalone, first class, interactive object. In considering design implications for these TTI models, our results also suggest the value in distinguishing between prompt artists---those users who embrace the constraint of creating images using only an input prompt---and practitioners, who may desire more fine-grained input and editing controls in comparison.

%% file: main.bbl

\begin{thebibliography}{62}


\ifx \showCODEN    \undefined \def \showCODEN     #1{\unskip}     \fi
\ifx \showDOI      \undefined \def \showDOI       #1{#1}\fi
\ifx \showISBNx    \undefined \def \showISBNx     #1{\unskip}     \fi
\ifx \showISBNxiii \undefined \def \showISBNxiii  #1{\unskip}     \fi
\ifx \showISSN     \undefined \def \showISSN      #1{\unskip}     \fi
\ifx \showLCCN     \undefined \def \showLCCN      #1{\unskip}     \fi
\ifx \shownote     \undefined \def \shownote      #1{#1}          \fi
\ifx \showarticletitle \undefined \def \showarticletitle #1{#1}   \fi
\ifx \showURL      \undefined \def \showURL       {\relax}        \fi
\providecommand\bibfield[2]{#2}
\providecommand\bibinfo[2]{#2}
\providecommand\natexlab[1]{#1}
\providecommand\showeprint[2][]{arXiv:#2}

\bibitem[\protect\citeauthoryear{??}{AIC}{[n.d.]}]%
        {AICAN73:online}
 \bibinfo{year}{[n.d.]}\natexlab{}.
\newblock \bibinfo{title}{AICAN}.
\newblock \bibinfo{howpublished}{\url{https://www.aican.io/}}.
\newblock
\newblock
\shownote{(Accessed on 08/31/2022).}


\bibitem[\protect\citeauthoryear{??}{Ima}{[n.d.]}]%
        {ImagenTe79:online}
 \bibinfo{year}{[n.d.]}\natexlab{}.
\newblock \bibinfo{title}{Imagen: Text-to-Image Diffusion Models}.
\newblock \bibinfo{howpublished}{\url{https://imagen.research.google/}}.
\newblock
\newblock
\shownote{(Accessed on 08/31/2022).}


\bibitem[\protect\citeauthoryear{??}{Mid}{[n.d.]}]%
        {Midjourn12:online}
 \bibinfo{year}{[n.d.]}\natexlab{}.
\newblock \bibinfo{title}{Midjourney}.
\newblock \bibinfo{howpublished}{\url{https://www.midjourney.com/home/}}.
\newblock
\newblock
\shownote{(Accessed on 09/01/2022).}


\bibitem[\protect\citeauthoryear{??}{Par}{[n.d.]}]%
        {PartiPat10:online}
 \bibinfo{year}{[n.d.]}\natexlab{}.
\newblock \bibinfo{title}{Parti: Pathways Autoregressive Text-to-Image Model}.
\newblock \bibinfo{howpublished}{\url{https://parti.research.google/}}.
\newblock
\newblock
\shownote{(Accessed on 08/31/2022).}


\bibitem[\protect\citeauthoryear{??}{Sta}{[n.d.]}]%
        {StableDi88:online}
 \bibinfo{year}{[n.d.]}\natexlab{}.
\newblock \bibinfo{title}{Stable Diffusion launch announcement —
  Stability.Ai}.
\newblock
  \bibinfo{howpublished}{\url{https://stability.ai/blog/stable-diffusion-announcement}}.
\newblock
\newblock
\shownote{(Accessed on 08/31/2022).}


\bibitem[\protect\citeauthoryear{Agüera~y Arcas}{Agüera~y Arcas}{2017}]%
        {aguera2017ArtInAI}
\bibfield{author}{\bibinfo{person}{Blaise Agüera~y Arcas}.}
  \bibinfo{year}{2017}\natexlab{}.
\newblock \showarticletitle{Art in the Age of Machine Intelligence}.
\newblock \bibinfo{journal}{\emph{Arts}} \bibinfo{volume}{6},
  \bibinfo{number}{4} (\bibinfo{year}{2017}).
\newblock
\showISSN{2076-0752}
\urldef\tempurl%
\url{https://doi.org/10.3390/arts6040018}
\showDOI{\tempurl}


\bibitem[\protect\citeauthoryear{Braun and Clarke}{Braun and Clarke}{2006}]%
        {braun2006using}
\bibfield{author}{\bibinfo{person}{Virginia Braun} {and}
  \bibinfo{person}{Victoria Clarke}.} \bibinfo{year}{2006}\natexlab{}.
\newblock \showarticletitle{Using thematic analysis in psychology}.
\newblock \bibinfo{journal}{\emph{Qualitative research in psychology}}
  \bibinfo{volume}{3}, \bibinfo{number}{2} (\bibinfo{year}{2006}),
  \bibinfo{pages}{77--101}.
\newblock


\bibitem[\protect\citeauthoryear{Brock, Donahue, and Simonyan}{Brock
  et~al\mbox{.}}{2018}]%
        {gan2018Brock}
\bibfield{author}{\bibinfo{person}{Andrew Brock}, \bibinfo{person}{Jeff
  Donahue}, {and} \bibinfo{person}{Karen Simonyan}.}
  \bibinfo{year}{2018}\natexlab{}.
\newblock \showarticletitle{Large Scale {GAN} Training for High Fidelity
  Natural Image Synthesis}.
\newblock \bibinfo{journal}{\emph{CoRR}}  \bibinfo{volume}{abs/1809.11096}
  (\bibinfo{year}{2018}).
\newblock
\showeprint[arXiv]{1809.11096}
\urldef\tempurl%
\url{http://arxiv.org/abs/1809.11096}
\showURL{%
\tempurl}


\bibitem[\protect\citeauthoryear{Browne}{Browne}{2022}]%
        {brown2022AIArtist}
\bibfield{author}{\bibinfo{person}{Kieran Browne}.}
  \bibinfo{year}{2022}\natexlab{}.
\newblock \showarticletitle{{Who (or What) Is an AI Artist?}}
\newblock \bibinfo{journal}{\emph{Leonardo}} \bibinfo{volume}{55},
  \bibinfo{number}{2} (\bibinfo{date}{04} \bibinfo{year}{2022}),
  \bibinfo{pages}{130--134}.
\newblock
\showISSN{0024-094X}
\urldef\tempurl%
\url{https://doi.org/10.1162/leon_a_02092}
\showDOI{\tempurl}
\showeprint{https://direct.mit.edu/leon/article-pdf/55/2/130/2004755/leon\_a\_02092.pdf}


\bibitem[\protect\citeauthoryear{Buschek, Mecke, Lehmann, and Dang}{Buschek
  et~al\mbox{.}}{2021}]%
        {buschek2021nine}
\bibfield{author}{\bibinfo{person}{Daniel Buschek}, \bibinfo{person}{Lukas
  Mecke}, \bibinfo{person}{Florian Lehmann}, {and} \bibinfo{person}{Hai Dang}.}
  \bibinfo{year}{2021}\natexlab{}.
\newblock \showarticletitle{Nine Potential Pitfalls when Designing Human-AI
  Co-Creative Systems}.
\newblock \bibinfo{journal}{\emph{Workshops at the International Conference on
  Intelligent User Interfaces (IUI)}} (\bibinfo{year}{2021}).
\newblock


\bibitem[\protect\citeauthoryear{Cetinic and She}{Cetinic and She}{2022}]%
        {aiart2022Cetinic}
\bibfield{author}{\bibinfo{person}{Eva Cetinic} {and} \bibinfo{person}{James
  She}.} \bibinfo{year}{2022}\natexlab{}.
\newblock \showarticletitle{Understanding and Creating Art with AI: Review and
  Outlook}.
\newblock \bibinfo{journal}{\emph{ACM Trans. Multimedia Comput. Commun. Appl.}}
  \bibinfo{volume}{18}, \bibinfo{number}{2}, Article \bibinfo{articleno}{66}
  (\bibinfo{date}{Feb} \bibinfo{year}{2022}), \bibinfo{numpages}{22}~pages.
\newblock
\showISSN{1551-6857}
\urldef\tempurl%
\url{https://doi.org/10.1145/3475799}
\showDOI{\tempurl}


\bibitem[\protect\citeauthoryear{Ch'ng}{Ch'ng}{2019}]%
        {10.1145/3326338}
\bibfield{author}{\bibinfo{person}{Eugene Ch'ng}.}
  \bibinfo{year}{2019}\natexlab{}.
\newblock \showarticletitle{Art by Computing Machinery: Is Machine Art
  Acceptable in the Artworld?}
\newblock \bibinfo{journal}{\emph{ACM Trans. Multimedia Comput. Commun. Appl.}}
  \bibinfo{volume}{15}, \bibinfo{number}{2s}, Article \bibinfo{articleno}{59}
  (\bibinfo{date}{Jul} \bibinfo{year}{2019}), \bibinfo{numpages}{17}~pages.
\newblock
\showISSN{1551-6857}
\urldef\tempurl%
\url{https://doi.org/10.1145/3326338}
\showDOI{\tempurl}


\bibitem[\protect\citeauthoryear{Christie's}{Christie's}{2018}]%
        {Thefirst77:online}
\bibfield{author}{\bibinfo{person}{Christie's}.}
  \bibinfo{year}{2018}\natexlab{}.
\newblock \bibinfo{title}{The first piece of AI-generated art to come to
  auction | Christie's}.
\newblock
  \bibinfo{howpublished}{\url{https://www.christies.com/features/a-collaboration-between-two-artists-one-human-one-a-machine-9332-1.aspx}}.
\newblock
\newblock
\shownote{(Accessed on 09/06/2022).}


\bibitem[\protect\citeauthoryear{Chung, Chang, and Adar}{Chung
  et~al\mbox{.}}{2021a}]%
        {chung2021gestural}
\bibfield{author}{\bibinfo{person}{John Joon~Young Chung},
  \bibinfo{person}{Minsuk Chang}, {and} \bibinfo{person}{Eytan Adar}.}
  \bibinfo{year}{2021}\natexlab{a}.
\newblock \showarticletitle{Gestural Inputs as Control Interaction for
  Generative Human-AI Co-Creation}.
\newblock \bibinfo{journal}{\emph{Workshops at the International Conference on
  Intelligent User Interfaces (IUI)}} (\bibinfo{year}{2021}).
\newblock


\bibitem[\protect\citeauthoryear{Chung, He, and Adar}{Chung
  et~al\mbox{.}}{2021b}]%
        {chung2021Intersection}
\bibfield{author}{\bibinfo{person}{John Joon~Young Chung},
  \bibinfo{person}{Shiqing He}, {and} \bibinfo{person}{Eytan Adar}.}
  \bibinfo{year}{2021}\natexlab{b}.
\newblock \showarticletitle{The Intersection of Users, Roles, Interactions, and
  Technologies in Creativity Support Tools}. In
  \bibinfo{booktitle}{\emph{Designing Interactive Systems Conference 2021}}
  \emph{(\bibinfo{series}{DIS '21})}. \bibinfo{publisher}{Association for
  Computing Machinery}, \bibinfo{address}{New York, NY, USA},
  \bibinfo{pages}{1817–1833}.
\newblock
\showISBNx{9781450384766}
\urldef\tempurl%
\url{https://doi.org/10.1145/3461778.3462050}
\showDOI{\tempurl}


\bibitem[\protect\citeauthoryear{Davis, Hsiao, Singh, Li, Moningi, and
  Magerko}{Davis et~al\mbox{.}}{2015}]%
        {davis2015Drawing}
\bibfield{author}{\bibinfo{person}{Nicholas Davis}, \bibinfo{person}{Chih-PIn
  Hsiao}, \bibinfo{person}{Kunwar~Yashraj Singh}, \bibinfo{person}{Lisa Li},
  \bibinfo{person}{Sanat Moningi}, {and} \bibinfo{person}{Brian Magerko}.}
  \bibinfo{year}{2015}\natexlab{}.
\newblock \showarticletitle{Drawing Apprentice: An Enactive Co-Creative Agent
  for Artistic Collaboration}. In \bibinfo{booktitle}{\emph{Proceedings of the
  2015 ACM SIGCHI Conference on Creativity and Cognition}}
  \emph{(\bibinfo{series}{C\&C '15})}. \bibinfo{publisher}{Association for
  Computing Machinery}, \bibinfo{address}{New York, NY, USA},
  \bibinfo{pages}{185–186}.
\newblock
\showISBNx{9781450335980}
\urldef\tempurl%
\url{https://doi.org/10.1145/2757226.2764555}
\showDOI{\tempurl}


\bibitem[\protect\citeauthoryear{Davis, Hsiao, Yashraj~Singh, Li, and
  Magerko}{Davis et~al\mbox{.}}{2016}]%
        {davis2016Drawing}
\bibfield{author}{\bibinfo{person}{Nicholas Davis}, \bibinfo{person}{Chih-PIn
  Hsiao}, \bibinfo{person}{Kunwar Yashraj~Singh}, \bibinfo{person}{Lisa Li},
  {and} \bibinfo{person}{Brian Magerko}.} \bibinfo{year}{2016}\natexlab{}.
\newblock \showarticletitle{Empirically Studying Participatory Sense-Making in
  Abstract Drawing with a Co-Creative Cognitive Agent}. In
  \bibinfo{booktitle}{\emph{Proceedings of the 21st International Conference on
  Intelligent User Interfaces}} \emph{(\bibinfo{series}{IUI '16})}.
  \bibinfo{publisher}{Association for Computing Machinery},
  \bibinfo{address}{New York, NY, USA}, \bibinfo{pages}{196–207}.
\newblock
\showISBNx{9781450341370}
\urldef\tempurl%
\url{https://doi.org/10.1145/2856767.2856795}
\showDOI{\tempurl}


\bibitem[\protect\citeauthoryear{Devlin, Chang, Lee, and Toutanova}{Devlin
  et~al\mbox{.}}{2019}]%
        {devlin2018Bert}
\bibfield{author}{\bibinfo{person}{Jacob Devlin}, \bibinfo{person}{Ming-Wei
  Chang}, \bibinfo{person}{Kenton Lee}, {and} \bibinfo{person}{Kristina
  Toutanova}.} \bibinfo{year}{2019}\natexlab{}.
\newblock \showarticletitle{{BERT}: Pre-training of Deep Bidirectional
  Transformers for Language Understanding}. In
  \bibinfo{booktitle}{\emph{Proceedings of the 2019 Conference of the North
  {A}merican Chapter of the Association for Computational Linguistics: Human
  Language Technologies, Volume 1 (Long and Short Papers)}}.
  \bibinfo{publisher}{Association for Computational Linguistics},
  \bibinfo{address}{Minneapolis, Minnesota}, \bibinfo{pages}{4171--4186}.
\newblock
\urldef\tempurl%
\url{https://doi.org/10.18653/v1/N19-1423}
\showDOI{\tempurl}


\bibitem[\protect\citeauthoryear{Dhariwal and Nichol}{Dhariwal and
  Nichol}{2021}]%
        {NEURIPS2021_49ad23d1}
\bibfield{author}{\bibinfo{person}{Prafulla Dhariwal} {and}
  \bibinfo{person}{Alexander Nichol}.} \bibinfo{year}{2021}\natexlab{}.
\newblock \showarticletitle{Diffusion Models Beat GANs on Image Synthesis}. In
  \bibinfo{booktitle}{\emph{Advances in Neural Information Processing
  Systems}}, \bibfield{editor}{\bibinfo{person}{M.~Ranzato},
  \bibinfo{person}{A.~Beygelzimer}, \bibinfo{person}{Y.~Dauphin},
  \bibinfo{person}{P.S. Liang}, {and} \bibinfo{person}{J.~Wortman Vaughan}}
  (Eds.), Vol.~\bibinfo{volume}{34}. \bibinfo{publisher}{Curran Associates,
  Inc.}, \bibinfo{pages}{8780--8794}.
\newblock
\urldef\tempurl%
\url{https://proceedings.neurips.cc/paper/2021/file/49ad23d1ec9fa4bd8d77d02681df5cfa-Paper.pdf}
\showURL{%
\tempurl}


\bibitem[\protect\citeauthoryear{Elgammal, Liu, Elhoseiny, and
  Mazzone}{Elgammal et~al\mbox{.}}{2017}]%
        {aican2017Elgammal}
\bibfield{author}{\bibinfo{person}{Ahmed Elgammal}, \bibinfo{person}{Bingchen
  Liu}, \bibinfo{person}{Mohamed Elhoseiny}, {and} \bibinfo{person}{Marian
  Mazzone}.} \bibinfo{year}{2017}\natexlab{}.
\newblock \bibinfo{title}{CAN: Creative Adversarial Networks, Generating "Art"
  by Learning About Styles and Deviating from Style Norms}.
\newblock
\newblock
\urldef\tempurl%
\url{https://doi.org/10.48550/ARXIV.1706.07068}
\showDOI{\tempurl}


\bibitem[\protect\citeauthoryear{Frich, MacDonald~Vermeulen, Remy, Biskjaer,
  and Dalsgaard}{Frich et~al\mbox{.}}{2019}]%
        {frih2019CSTinHCI}
\bibfield{author}{\bibinfo{person}{Jonas Frich}, \bibinfo{person}{Lindsay
  MacDonald~Vermeulen}, \bibinfo{person}{Christian Remy},
  \bibinfo{person}{Michael~Mose Biskjaer}, {and} \bibinfo{person}{Peter
  Dalsgaard}.} \bibinfo{year}{2019}\natexlab{}.
\newblock \showarticletitle{Mapping the Landscape of Creativity Support Tools
  in HCI}. In \bibinfo{booktitle}{\emph{Proceedings of the 2019 CHI Conference
  on Human Factors in Computing Systems}} \emph{(\bibinfo{series}{CHI '19})}.
  \bibinfo{publisher}{Association for Computing Machinery},
  \bibinfo{address}{New York, NY, USA}, \bibinfo{pages}{1–18}.
\newblock
\showISBNx{9781450359702}
\urldef\tempurl%
\url{https://doi.org/10.1145/3290605.3300619}
\showDOI{\tempurl}


\bibitem[\protect\citeauthoryear{Gatys, Ecker, and Bethge}{Gatys
  et~al\mbox{.}}{2015}]%
        {gatys2015NeuralStyle}
\bibfield{author}{\bibinfo{person}{Leon~A. Gatys},
  \bibinfo{person}{Alexander~S. Ecker}, {and} \bibinfo{person}{Matthias
  Bethge}.} \bibinfo{year}{2015}\natexlab{}.
\newblock \bibinfo{title}{A Neural Algorithm of Artistic Style}.
\newblock
\newblock
\urldef\tempurl%
\url{https://doi.org/10.48550/ARXIV.1508.06576}
\showDOI{\tempurl}


\bibitem[\protect\citeauthoryear{Goodfellow, Pouget-Abadie, Mirza, Xu,
  Warde-Farley, Ozair, Courville, and Bengio}{Goodfellow et~al\mbox{.}}{2020}]%
        {gan2014Goodfellow}
\bibfield{author}{\bibinfo{person}{Ian Goodfellow}, \bibinfo{person}{Jean
  Pouget-Abadie}, \bibinfo{person}{Mehdi Mirza}, \bibinfo{person}{Bing Xu},
  \bibinfo{person}{David Warde-Farley}, \bibinfo{person}{Sherjil Ozair},
  \bibinfo{person}{Aaron Courville}, {and} \bibinfo{person}{Yoshua Bengio}.}
  \bibinfo{year}{2020}\natexlab{}.
\newblock \showarticletitle{Generative Adversarial Networks}.
\newblock \bibinfo{journal}{\emph{Commun. ACM}} \bibinfo{volume}{63},
  \bibinfo{number}{11} (\bibinfo{date}{Oct} \bibinfo{year}{2020}),
  \bibinfo{pages}{139–144}.
\newblock
\showISSN{0001-0782}
\urldef\tempurl%
\url{https://doi.org/10.1145/3422622}
\showDOI{\tempurl}


\bibitem[\protect\citeauthoryear{Grabe, Gonz{\'a}lez-Duque, Risi, and
  Zhu}{Grabe et~al\mbox{.}}{2022}]%
        {grabe2022towards}
\bibfield{author}{\bibinfo{person}{Imke Grabe}, \bibinfo{person}{Miguel
  Gonz{\'a}lez-Duque}, \bibinfo{person}{Sebastian Risi}, {and}
  \bibinfo{person}{Jichen Zhu}.} \bibinfo{year}{2022}\natexlab{}.
\newblock \showarticletitle{Towards a Framework for Human-AI Interaction
  Patterns in Co-Creative GAN Applications}.
\newblock \bibinfo{journal}{\emph{Workshops at the International Conference on
  Intelligent User Interfaces (IUI)}} (\bibinfo{year}{2022}).
\newblock


\bibitem[\protect\citeauthoryear{Grba}{Grba}{2019}]%
        {grba2019forensics}
\bibfield{author}{\bibinfo{person}{Dejan Grba}.}
  \bibinfo{year}{2019}\natexlab{}.
\newblock \showarticletitle{Forensics of a molten crystal: challenges of
  archiving and representing contemporary generative art}.
\newblock \bibinfo{journal}{\emph{ISSUE Annual Art Journal: Erase}}
  \bibinfo{volume}{8}, \bibinfo{number}{3-15} (\bibinfo{year}{2019}),
  \bibinfo{pages}{5}.
\newblock


\bibitem[\protect\citeauthoryear{Grba}{Grba}{2021}]%
        {grba2021brittle}
\bibfield{author}{\bibinfo{person}{Dejan Grba}.}
  \bibinfo{year}{2021}\natexlab{}.
\newblock \showarticletitle{Brittle Opacity: Ambiguities of the Creative AI}.
  In \bibinfo{booktitle}{\emph{Proceedings of the xCoAx, 9th Conference on
  Computation, Communication, Aesthetics \& X Proceedings, xCoAx, Graz,
  Austria}}. \bibinfo{pages}{12--16}.
\newblock


\bibitem[\protect\citeauthoryear{Grba}{Grba}{2022}]%
        {grba2022DeepElse}
\bibfield{author}{\bibinfo{person}{Dejan Grba}.}
  \bibinfo{year}{2022}\natexlab{}.
\newblock \showarticletitle{Deep Else: A Critical Framework for AI Art}.
\newblock \bibinfo{journal}{\emph{Digital}} \bibinfo{volume}{2},
  \bibinfo{number}{1} (\bibinfo{year}{2022}), \bibinfo{pages}{1--32}.
\newblock
\showISSN{2673-6470}
\urldef\tempurl%
\url{https://doi.org/10.3390/digital2010001}
\showDOI{\tempurl}


\bibitem[\protect\citeauthoryear{Hertzmann}{Hertzmann}{2018}]%
        {hertzmann2018ComputersArt}
\bibfield{author}{\bibinfo{person}{Aaron Hertzmann}.}
  \bibinfo{year}{2018}\natexlab{}.
\newblock \showarticletitle{Can Computers Create Art?}
\newblock \bibinfo{journal}{\emph{Arts}} \bibinfo{volume}{7},
  \bibinfo{number}{2} (\bibinfo{year}{2018}).
\newblock
\showISSN{2076-0752}
\urldef\tempurl%
\url{https://doi.org/10.3390/arts7020018}
\showDOI{\tempurl}


\bibitem[\protect\citeauthoryear{Hertzmann}{Hertzmann}{2020}]%
        {hertzmann2020ComputersArt}
\bibfield{author}{\bibinfo{person}{Aaron Hertzmann}.}
  \bibinfo{year}{2020}\natexlab{}.
\newblock \showarticletitle{Computers Do Not Make Art, People Do}.
\newblock \bibinfo{journal}{\emph{Commun. ACM}} \bibinfo{volume}{63},
  \bibinfo{number}{5} (\bibinfo{date}{Apr} \bibinfo{year}{2020}),
  \bibinfo{pages}{45–48}.
\newblock
\showISSN{0001-0782}
\urldef\tempurl%
\url{https://doi.org/10.1145/3347092}
\showDOI{\tempurl}


\bibitem[\protect\citeauthoryear{Ho, Saharia, Chan, Fleet, Norouzi, and
  Salimans}{Ho et~al\mbox{.}}{2022}]%
        {ho2022cascaded}
\bibfield{author}{\bibinfo{person}{Jonathan Ho}, \bibinfo{person}{Chitwan
  Saharia}, \bibinfo{person}{William Chan}, \bibinfo{person}{David~J Fleet},
  \bibinfo{person}{Mohammad Norouzi}, {and} \bibinfo{person}{Tim Salimans}.}
  \bibinfo{year}{2022}\natexlab{}.
\newblock \showarticletitle{Cascaded Diffusion Models for High Fidelity Image
  Generation.}
\newblock \bibinfo{journal}{\emph{J. Mach. Learn. Res.}}  \bibinfo{volume}{23}
  (\bibinfo{year}{2022}), \bibinfo{pages}{47--1}.
\newblock


\bibitem[\protect\citeauthoryear{Hodhod and Magerko}{Hodhod and
  Magerko}{2016}]%
        {hodhod2016Storytelling}
\bibfield{author}{\bibinfo{person}{Rania Hodhod} {and} \bibinfo{person}{Brian
  Magerko}.} \bibinfo{year}{2016}\natexlab{}.
\newblock \showarticletitle{Closing the Cognitive Gap between Humans and
  Interactive Narrative Agents Using Shared Mental Models}. In
  \bibinfo{booktitle}{\emph{Proceedings of the 21st International Conference on
  Intelligent User Interfaces}} \emph{(\bibinfo{series}{IUI '16})}.
  \bibinfo{publisher}{Association for Computing Machinery},
  \bibinfo{address}{New York, NY, USA}, \bibinfo{pages}{135–146}.
\newblock
\showISBNx{9781450341370}
\urldef\tempurl%
\url{https://doi.org/10.1145/2856767.2856774}
\showDOI{\tempurl}


\bibitem[\protect\citeauthoryear{Huang, Koops, Newton-Rex, Dinculescu, and
  Cai}{Huang et~al\mbox{.}}{2020}]%
        {huang2020AISong}
\bibfield{author}{\bibinfo{person}{Cheng-Zhi~Anna Huang},
  \bibinfo{person}{Hendrik~Vincent Koops}, \bibinfo{person}{Ed Newton-Rex},
  \bibinfo{person}{Monica Dinculescu}, {and} \bibinfo{person}{Carrie~J. Cai}.}
  \bibinfo{year}{2020}\natexlab{}.
\newblock \showarticletitle{AI Song Contest: Human-AI Co-Creation in
  Songwriting}.
\newblock  (\bibinfo{year}{2020}).
\newblock
\urldef\tempurl%
\url{https://doi.org/10.48550/ARXIV.2010.05388}
\showDOI{\tempurl}


\bibitem[\protect\citeauthoryear{Hwang}{Hwang}{2022}]%
        {hwang2022Creative}
\bibfield{author}{\bibinfo{person}{Angel Hsing-Chi Hwang}.}
  \bibinfo{year}{2022}\natexlab{}.
\newblock \showarticletitle{Too Late to Be Creative? AI-Empowered Tools in
  Creative Processes}. In \bibinfo{booktitle}{\emph{Extended Abstracts of the
  2022 CHI Conference on Human Factors in Computing Systems}}
  \emph{(\bibinfo{series}{CHI EA '22})}. \bibinfo{publisher}{Association for
  Computing Machinery}, \bibinfo{address}{New York, NY, USA}, Article
  \bibinfo{articleno}{38}, \bibinfo{numpages}{9}~pages.
\newblock
\showISBNx{9781450391566}
\urldef\tempurl%
\url{https://doi.org/10.1145/3491101.3503549}
\showDOI{\tempurl}


\bibitem[\protect\citeauthoryear{Jeon, Jin, Shih, and Han}{Jeon
  et~al\mbox{.}}{2021}]%
        {jeon2021Fashion}
\bibfield{author}{\bibinfo{person}{Youngseung Jeon}, \bibinfo{person}{Seungwan
  Jin}, \bibinfo{person}{Patrick~C. Shih}, {and} \bibinfo{person}{Kyungsik
  Han}.} \bibinfo{year}{2021}\natexlab{}.
\newblock \showarticletitle{FashionQ: An AI-Driven Creativity Support Tool for
  Facilitating Ideation in Fashion Design}. In
  \bibinfo{booktitle}{\emph{Proceedings of the 2021 CHI Conference on Human
  Factors in Computing Systems}} \emph{(\bibinfo{series}{CHI '21})}.
  \bibinfo{publisher}{Association for Computing Machinery},
  \bibinfo{address}{New York, NY, USA}, Article \bibinfo{articleno}{576},
  \bibinfo{numpages}{18}~pages.
\newblock
\showISBNx{9781450380966}
\urldef\tempurl%
\url{https://doi.org/10.1145/3411764.3445093}
\showDOI{\tempurl}


\bibitem[\protect\citeauthoryear{Jing, Yang, Feng, Ye, Yu, and Song}{Jing
  et~al\mbox{.}}{2020}]%
        {jing2020NeuralStyle}
\bibfield{author}{\bibinfo{person}{Yongcheng Jing}, \bibinfo{person}{Yezhou
  Yang}, \bibinfo{person}{Zunlei Feng}, \bibinfo{person}{Jingwen Ye},
  \bibinfo{person}{Yizhou Yu}, {and} \bibinfo{person}{Mingli Song}.}
  \bibinfo{year}{2020}\natexlab{}.
\newblock \showarticletitle{Neural Style Transfer: A Review}.
\newblock \bibinfo{journal}{\emph{IEEE Transactions on Visualization and
  Computer Graphics}} \bibinfo{volume}{26}, \bibinfo{number}{11}
  (\bibinfo{year}{2020}), \bibinfo{pages}{3365--3385}.
\newblock
\urldef\tempurl%
\url{https://doi.org/10.1109/TVCG.2019.2921336}
\showDOI{\tempurl}


\bibitem[\protect\citeauthoryear{Karimi, Davis, Maher, Grace, and Lee}{Karimi
  et~al\mbox{.}}{2019}]%
        {karimi2019Drawing}
\bibfield{author}{\bibinfo{person}{Pegah Karimi}, \bibinfo{person}{Nicholas
  Davis}, \bibinfo{person}{Mary~Lou Maher}, \bibinfo{person}{Kazjon Grace},
  {and} \bibinfo{person}{Lina Lee}.} \bibinfo{year}{2019}\natexlab{}.
\newblock \showarticletitle{Relating Cognitive Models of Design Creativity to
  the Similarity of Sketches Generated by an AI Partner}. In
  \bibinfo{booktitle}{\emph{Proceedings of the 2019 on Creativity and
  Cognition}} \emph{(\bibinfo{series}{C\&C '19})}.
  \bibinfo{publisher}{Association for Computing Machinery},
  \bibinfo{address}{New York, NY, USA}, \bibinfo{pages}{259–270}.
\newblock
\showISBNx{9781450359177}
\urldef\tempurl%
\url{https://doi.org/10.1145/3325480.3325488}
\showDOI{\tempurl}


\bibitem[\protect\citeauthoryear{Karras, Laine, and Aila}{Karras
  et~al\mbox{.}}{2019}]%
        {Karras_2019_CVPR}
\bibfield{author}{\bibinfo{person}{Tero Karras}, \bibinfo{person}{Samuli
  Laine}, {and} \bibinfo{person}{Timo Aila}.} \bibinfo{year}{2019}\natexlab{}.
\newblock \showarticletitle{A Style-Based Generator Architecture for Generative
  Adversarial Networks}. In \bibinfo{booktitle}{\emph{Proceedings of the
  IEEE/CVF Conference on Computer Vision and Pattern Recognition (CVPR)}}.
\newblock


\bibitem[\protect\citeauthoryear{Louie, Coenen, Huang, Terry, and Cai}{Louie
  et~al\mbox{.}}{2020}]%
        {louie2020Music}
\bibfield{author}{\bibinfo{person}{Ryan Louie}, \bibinfo{person}{Andy Coenen},
  \bibinfo{person}{Cheng~Zhi Huang}, \bibinfo{person}{Michael Terry}, {and}
  \bibinfo{person}{Carrie~J. Cai}.} \bibinfo{year}{2020}\natexlab{}.
\newblock \showarticletitle{Novice-AI Music Co-Creation via AI-Steering Tools
  for Deep Generative Models}. In \bibinfo{booktitle}{\emph{Proceedings of the
  2020 CHI Conference on Human Factors in Computing Systems}}
  \emph{(\bibinfo{series}{CHI '20})}. \bibinfo{publisher}{Association for
  Computing Machinery}, \bibinfo{address}{New York, NY, USA},
  \bibinfo{pages}{1–13}.
\newblock
\showISBNx{9781450367080}
\urldef\tempurl%
\url{https://doi.org/10.1145/3313831.3376739}
\showDOI{\tempurl}


\bibitem[\protect\citeauthoryear{Mansimov, Parisotto, Ba, and
  Salakhutdinov}{Mansimov et~al\mbox{.}}{2015}]%
        {mansimov2015ImgFromCaption}
\bibfield{author}{\bibinfo{person}{Elman Mansimov}, \bibinfo{person}{Emilio
  Parisotto}, \bibinfo{person}{Jimmy~Lei Ba}, {and} \bibinfo{person}{Ruslan
  Salakhutdinov}.} \bibinfo{year}{2015}\natexlab{}.
\newblock \bibinfo{title}{Generating Images from Captions with Attention}.
\newblock
\newblock
\urldef\tempurl%
\url{https://doi.org/10.48550/ARXIV.1511.02793}
\showDOI{\tempurl}


\bibitem[\protect\citeauthoryear{Mazzone and Elgammal}{Mazzone and
  Elgammal}{2019}]%
        {mazzone2019ArtCreativityAI}
\bibfield{author}{\bibinfo{person}{Marian Mazzone} {and} \bibinfo{person}{Ahmed
  Elgammal}.} \bibinfo{year}{2019}\natexlab{}.
\newblock \showarticletitle{Art, Creativity, and the Potential of Artificial
  Intelligence}.
\newblock \bibinfo{journal}{\emph{Arts}} \bibinfo{volume}{8},
  \bibinfo{number}{1} (\bibinfo{year}{2019}).
\newblock
\showISSN{2076-0752}
\urldef\tempurl%
\url{https://doi.org/10.3390/arts8010026}
\showDOI{\tempurl}


\bibitem[\protect\citeauthoryear{McCormack, Gifford, Hutchings,
  Llano~Rodriguez, Yee-King, and d'Inverno}{McCormack et~al\mbox{.}}{2019}]%
        {mccormack2019Music}
\bibfield{author}{\bibinfo{person}{Jon McCormack}, \bibinfo{person}{Toby
  Gifford}, \bibinfo{person}{Patrick Hutchings}, \bibinfo{person}{Maria~Teresa
  Llano~Rodriguez}, \bibinfo{person}{Matthew Yee-King}, {and}
  \bibinfo{person}{Mark d'Inverno}.} \bibinfo{year}{2019}\natexlab{}.
\newblock \showarticletitle{In a Silent Way: Communication Between AI and
  Improvising Musicians Beyond Sound}. In \bibinfo{booktitle}{\emph{Proceedings
  of the 2019 CHI Conference on Human Factors in Computing Systems}}
  \emph{(\bibinfo{series}{CHI '19})}. \bibinfo{publisher}{Association for
  Computing Machinery}, \bibinfo{address}{New York, NY, USA},
  \bibinfo{pages}{1–11}.
\newblock
\showISBNx{9781450359702}
\urldef\tempurl%
\url{https://doi.org/10.1145/3290605.3300268}
\showDOI{\tempurl}


\bibitem[\protect\citeauthoryear{McFarland}{McFarland}{2016}]%
        {GoogleDeepDream81:online}
\bibfield{author}{\bibinfo{person}{Matt McFarland}.}
  \bibinfo{year}{2016}\natexlab{}.
\newblock \bibinfo{title}{Google’s psychedelic ‘paint brush’ raises the
  oldest question in art - The Washington Post}.
\newblock
  \bibinfo{howpublished}{\url{https://www.washingtonpost.com/news/innovations/wp/2016/03/10/googles-psychedelic-paint-brush-raises-the-oldest-question-in-art/}}.
\newblock
\newblock
\shownote{(Accessed on 09/14/2022).}


\bibitem[\protect\citeauthoryear{Miles and Huberman}{Miles and
  Huberman}{1984}]%
        {miles1984drawing}
\bibfield{author}{\bibinfo{person}{Matthew~B Miles} {and}
  \bibinfo{person}{A~Michael Huberman}.} \bibinfo{year}{1984}\natexlab{}.
\newblock \showarticletitle{Drawing valid meaning from qualitative data: Toward
  a shared craft}.
\newblock \bibinfo{journal}{\emph{Educational researcher}}
  \bibinfo{volume}{13}, \bibinfo{number}{5} (\bibinfo{year}{1984}),
  \bibinfo{pages}{20--30}.
\newblock


\bibitem[\protect\citeauthoryear{Mordvintsev, Olah, and Tyka}{Mordvintsev
  et~al\mbox{.}}{2015}]%
        {mordvintsev2015DeepDreams}
\bibfield{author}{\bibinfo{person}{Alexander Mordvintsev},
  \bibinfo{person}{Christopher Olah}, {and} \bibinfo{person}{Mike Tyka}.}
  \bibinfo{year}{2015}\natexlab{}.
\newblock \bibinfo{title}{Inceptionism: Going Deeper into Neural Networks}.
\newblock
\newblock
\urldef\tempurl%
\url{https://research.googleblog.com/2015/06/inceptionism-going-deeper-into-neural.html}
\showURL{%
\tempurl}


\bibitem[\protect\citeauthoryear{Muller, Weisz, and Geyer}{Muller
  et~al\mbox{.}}{2020}]%
        {muller2020mixed}
\bibfield{author}{\bibinfo{person}{Michael Muller}, \bibinfo{person}{Justin~D
  Weisz}, {and} \bibinfo{person}{Werner Geyer}.}
  \bibinfo{year}{2020}\natexlab{}.
\newblock \showarticletitle{Mixed Initiative Generative AI Interfaces: An
  Analytic Framework for Generative AI Applications}. In
  \bibinfo{booktitle}{\emph{Proceedings of the Workshop The Future of
  Co-Creative Systems-A Workshop on Human-Computer Co-Creativity of the 11th
  International Conference on Computational Creativity (ICCC 2020)}}.
\newblock


\bibitem[\protect\citeauthoryear{Oh, Song, Choi, Kim, Lee, and Suh}{Oh
  et~al\mbox{.}}{2018}]%
        {oh2018Drawing}
\bibfield{author}{\bibinfo{person}{Changhoon Oh}, \bibinfo{person}{Jungwoo
  Song}, \bibinfo{person}{Jinhan Choi}, \bibinfo{person}{Seonghyeon Kim},
  \bibinfo{person}{Sungwoo Lee}, {and} \bibinfo{person}{Bongwon Suh}.}
  \bibinfo{year}{2018}\natexlab{}.
\newblock \showarticletitle{I Lead, You Help but Only with Enough Details:
  Understanding User Experience of Co-Creation with Artificial Intelligence}.
  In \bibinfo{booktitle}{\emph{Proceedings of the 2018 CHI Conference on Human
  Factors in Computing Systems}} \emph{(\bibinfo{series}{CHI '18})}.
  \bibinfo{publisher}{Association for Computing Machinery},
  \bibinfo{address}{New York, NY, USA}, \bibinfo{pages}{1–13}.
\newblock
\showISBNx{9781450356206}
\urldef\tempurl%
\url{https://doi.org/10.1145/3173574.3174223}
\showDOI{\tempurl}


\bibitem[\protect\citeauthoryear{OpenAI}{OpenAI}{[n.d.]a}]%
        {DALLE279:online}
\bibfield{author}{\bibinfo{person}{OpenAI}.}
  \bibinfo{year}{[n.d.]}\natexlab{a}.
\newblock \bibinfo{title}{DALL·E 2}.
\newblock \bibinfo{howpublished}{\url{https://openai.com/dall-e-2/}}.
\newblock
\newblock
\shownote{(Accessed on 08/31/2022).}


\bibitem[\protect\citeauthoryear{OpenAI}{OpenAI}{[n.d.]b}]%
        {DALLECr11:online}
\bibfield{author}{\bibinfo{person}{OpenAI}.}
  \bibinfo{year}{[n.d.]}\natexlab{b}.
\newblock \bibinfo{title}{DALL·E: Creating Images from Text}.
\newblock \bibinfo{howpublished}{\url{https://openai.com/blog/dall-e/}}.
\newblock
\newblock
\shownote{(Accessed on 08/31/2022).}


\bibitem[\protect\citeauthoryear{Perrone and Edwards}{Perrone and
  Edwards}{2019}]%
        {perone2019Chatbot}
\bibfield{author}{\bibinfo{person}{Allison Perrone} {and}
  \bibinfo{person}{Justin Edwards}.} \bibinfo{year}{2019}\natexlab{}.
\newblock \showarticletitle{Chatbots as Unwitting Actors}. In
  \bibinfo{booktitle}{\emph{Proceedings of the 1st International Conference on
  Conversational User Interfaces}} \emph{(\bibinfo{series}{CUI '19})}.
  \bibinfo{publisher}{Association for Computing Machinery},
  \bibinfo{address}{New York, NY, USA}, Article \bibinfo{articleno}{2},
  \bibinfo{numpages}{2}~pages.
\newblock
\showISBNx{9781450371872}
\urldef\tempurl%
\url{https://doi.org/10.1145/3342775.3342799}
\showDOI{\tempurl}


\bibitem[\protect\citeauthoryear{Quanz, Sun, Deshpande, Shah, and Park}{Quanz
  et~al\mbox{.}}{2020}]%
        {quanz2020machine}
\bibfield{author}{\bibinfo{person}{Brian Quanz}, \bibinfo{person}{Wei Sun},
  \bibinfo{person}{Ajay Deshpande}, \bibinfo{person}{Dhruv Shah}, {and}
  \bibinfo{person}{Jae-eun Park}.} \bibinfo{year}{2020}\natexlab{}.
\newblock \showarticletitle{Machine learning based co-creative design
  framework}.
\newblock \bibinfo{journal}{\emph{arXiv preprint arXiv:2001.08791}}
  (\bibinfo{year}{2020}).
\newblock


\bibitem[\protect\citeauthoryear{Raffel, Shazeer, Roberts, Lee, Narang, Matena,
  Zhou, Li, Liu, et~al\mbox{.}}{Raffel et~al\mbox{.}}{2020}]%
        {raffel2020exploring}
\bibfield{author}{\bibinfo{person}{Colin Raffel}, \bibinfo{person}{Noam
  Shazeer}, \bibinfo{person}{Adam Roberts}, \bibinfo{person}{Katherine Lee},
  \bibinfo{person}{Sharan Narang}, \bibinfo{person}{Michael Matena},
  \bibinfo{person}{Yanqi Zhou}, \bibinfo{person}{Wei Li},
  \bibinfo{person}{Peter~J Liu}, {et~al\mbox{.}}}
  \bibinfo{year}{2020}\natexlab{}.
\newblock \showarticletitle{Exploring the limits of transfer learning with a
  unified text-to-text transformer.}
\newblock \bibinfo{journal}{\emph{J. Mach. Learn. Res.}} \bibinfo{volume}{21},
  \bibinfo{number}{140} (\bibinfo{year}{2020}), \bibinfo{pages}{1--67}.
\newblock


\bibitem[\protect\citeauthoryear{Ramesh, Dhariwal, Nichol, Chu, and
  Chen}{Ramesh et~al\mbox{.}}{2022}]%
        {ramesh2022DALLE2}
\bibfield{author}{\bibinfo{person}{Aditya Ramesh}, \bibinfo{person}{Prafulla
  Dhariwal}, \bibinfo{person}{Alex Nichol}, \bibinfo{person}{Casey Chu}, {and}
  \bibinfo{person}{Mark Chen}.} \bibinfo{year}{2022}\natexlab{}.
\newblock \bibinfo{title}{Hierarchical Text-Conditional Image Generation with
  CLIP Latents}.
\newblock
\newblock
\urldef\tempurl%
\url{https://doi.org/10.48550/ARXIV.2204.06125}
\showDOI{\tempurl}


\bibitem[\protect\citeauthoryear{Ramesh, Pavlov, Goh, Gray, Voss, Radford,
  Chen, and Sutskever}{Ramesh et~al\mbox{.}}{2021}]%
        {pmlr-v139-ramesh21a}
\bibfield{author}{\bibinfo{person}{Aditya Ramesh}, \bibinfo{person}{Mikhail
  Pavlov}, \bibinfo{person}{Gabriel Goh}, \bibinfo{person}{Scott Gray},
  \bibinfo{person}{Chelsea Voss}, \bibinfo{person}{Alec Radford},
  \bibinfo{person}{Mark Chen}, {and} \bibinfo{person}{Ilya Sutskever}.}
  \bibinfo{year}{2021}\natexlab{}.
\newblock \showarticletitle{Zero-Shot Text-to-Image Generation}. In
  \bibinfo{booktitle}{\emph{Proceedings of the 38th International Conference on
  Machine Learning}} \emph{(\bibinfo{series}{Proceedings of Machine Learning
  Research})}, \bibfield{editor}{\bibinfo{person}{Marina Meila} {and}
  \bibinfo{person}{Tong Zhang}} (Eds.), Vol.~\bibinfo{volume}{139}.
  \bibinfo{publisher}{PMLR}, \bibinfo{pages}{8821--8831}.
\newblock
\urldef\tempurl%
\url{https://proceedings.mlr.press/v139/ramesh21a.html}
\showURL{%
\tempurl}


\bibitem[\protect\citeauthoryear{Rombach, Blattmann, Lorenz, Esser, and
  Ommer}{Rombach et~al\mbox{.}}{2022}]%
        {Rombach_2022_CVPR}
\bibfield{author}{\bibinfo{person}{Robin Rombach}, \bibinfo{person}{Andreas
  Blattmann}, \bibinfo{person}{Dominik Lorenz}, \bibinfo{person}{Patrick
  Esser}, {and} \bibinfo{person}{Bj\"orn Ommer}.}
  \bibinfo{year}{2022}\natexlab{}.
\newblock \showarticletitle{High-Resolution Image Synthesis With Latent
  Diffusion Models}. In \bibinfo{booktitle}{\emph{Proceedings of the IEEE/CVF
  Conference on Computer Vision and Pattern Recognition (CVPR)}}.
  \bibinfo{pages}{10684--10695}.
\newblock


\bibitem[\protect\citeauthoryear{Saharia, Chan, Saxena, Li, Whang, Denton,
  Ghasemipour, Ayan, Mahdavi, Lopes, Salimans, Ho, Fleet, and Norouzi}{Saharia
  et~al\mbox{.}}{2022}]%
        {saharia2022Imagen}
\bibfield{author}{\bibinfo{person}{Chitwan Saharia}, \bibinfo{person}{William
  Chan}, \bibinfo{person}{Saurabh Saxena}, \bibinfo{person}{Lala Li},
  \bibinfo{person}{Jay Whang}, \bibinfo{person}{Emily Denton},
  \bibinfo{person}{Seyed Kamyar~Seyed Ghasemipour},
  \bibinfo{person}{Burcu~Karagol Ayan}, \bibinfo{person}{S.~Sara Mahdavi},
  \bibinfo{person}{Rapha~Gontijo Lopes}, \bibinfo{person}{Tim Salimans},
  \bibinfo{person}{Jonathan Ho}, \bibinfo{person}{David~J Fleet}, {and}
  \bibinfo{person}{Mohammad Norouzi}.} \bibinfo{year}{2022}\natexlab{}.
\newblock \bibinfo{title}{Photorealistic Text-to-Image Diffusion Models with
  Deep Language Understanding}.
\newblock
\newblock
\urldef\tempurl%
\url{https://doi.org/10.48550/ARXIV.2205.11487}
\showDOI{\tempurl}


\bibitem[\protect\citeauthoryear{Sbai, Elhoseiny, Bordes, LeCun, and
  Couprie}{Sbai et~al\mbox{.}}{2018}]%
        {Sbai_2018_ECCV_Workshops}
\bibfield{author}{\bibinfo{person}{Othman Sbai}, \bibinfo{person}{Mohamed
  Elhoseiny}, \bibinfo{person}{Antoine Bordes}, \bibinfo{person}{Yann LeCun},
  {and} \bibinfo{person}{Camille Couprie}.} \bibinfo{year}{2018}\natexlab{}.
\newblock \showarticletitle{DesIGN: Design Inspiration from Generative
  Networks}. In \bibinfo{booktitle}{\emph{Proceedings of the European
  Conference on Computer Vision (ECCV) Workshops}}.
\newblock


\bibitem[\protect\citeauthoryear{Shi, Cao, Ma, Chen, and Liu}{Shi
  et~al\mbox{.}}{2020}]%
        {shi2020Storyboarding}
\bibfield{author}{\bibinfo{person}{Yang Shi}, \bibinfo{person}{Nan Cao},
  \bibinfo{person}{Xiaojuan Ma}, \bibinfo{person}{Siji Chen}, {and}
  \bibinfo{person}{Pei Liu}.} \bibinfo{year}{2020}\natexlab{}.
\newblock \showarticletitle{EmoG: Supporting the Sketching of Emotional
  Expressions for Storyboarding}. In \bibinfo{booktitle}{\emph{Proceedings of
  the 2020 CHI Conference on Human Factors in Computing Systems}}
  \emph{(\bibinfo{series}{CHI '20})}. \bibinfo{publisher}{Association for
  Computing Machinery}, \bibinfo{address}{New York, NY, USA},
  \bibinfo{pages}{1–12}.
\newblock
\showISBNx{9781450367080}
\urldef\tempurl%
\url{https://doi.org/10.1145/3313831.3376520}
\showDOI{\tempurl}


\bibitem[\protect\citeauthoryear{Spoto and Oleynik}{Spoto and
  Oleynik}{[n.d.]}]%
        {spotoLibrary}
\bibfield{author}{\bibinfo{person}{Angie Spoto} {and} \bibinfo{person}{Natalia
  Oleynik}.} \bibinfo{year}{[n.d.]}\natexlab{}.
\newblock \bibinfo{title}{Library of Mixed-Initiative Creative Interfaces}.
\newblock
  \bibinfo{howpublished}{\url{http://mici.codingconduct.cc/aboutmicis/}}.
\newblock
\newblock
\shownote{(Accessed on 08/31/2022).}


\bibitem[\protect\citeauthoryear{Yannakakis, Liapis, and
  Alexopoulos}{Yannakakis et~al\mbox{.}}{2014}]%
        {yannakakis2014mixed}
\bibfield{author}{\bibinfo{person}{Georgios~N Yannakakis},
  \bibinfo{person}{Antonios Liapis}, {and} \bibinfo{person}{Constantine
  Alexopoulos}.} \bibinfo{year}{2014}\natexlab{}.
\newblock \showarticletitle{Mixed-initiative co-creativity}.
\newblock  (\bibinfo{year}{2014}).
\newblock


\bibitem[\protect\citeauthoryear{Yu, Xu, Koh, Luong, Baid, Wang, Vasudevan, Ku,
  Yang, Ayan, Hutchinson, Han, Parekh, Li, Zhang, Baldridge, and Wu}{Yu
  et~al\mbox{.}}{2022}]%
        {yu2022Parti}
\bibfield{author}{\bibinfo{person}{Jiahui Yu}, \bibinfo{person}{Yuanzhong Xu},
  \bibinfo{person}{Jing~Yu Koh}, \bibinfo{person}{Thang Luong},
  \bibinfo{person}{Gunjan Baid}, \bibinfo{person}{Zirui Wang},
  \bibinfo{person}{Vijay Vasudevan}, \bibinfo{person}{Alexander Ku},
  \bibinfo{person}{Yinfei Yang}, \bibinfo{person}{Burcu~Karagol Ayan},
  \bibinfo{person}{Ben Hutchinson}, \bibinfo{person}{Wei Han},
  \bibinfo{person}{Zarana Parekh}, \bibinfo{person}{Xin Li},
  \bibinfo{person}{Han Zhang}, \bibinfo{person}{Jason Baldridge}, {and}
  \bibinfo{person}{Yonghui Wu}.} \bibinfo{year}{2022}\natexlab{}.
\newblock \bibinfo{title}{Scaling Autoregressive Models for Content-Rich
  Text-to-Image Generation}.
\newblock
\newblock
\urldef\tempurl%
\url{https://doi.org/10.48550/ARXIV.2206.10789}
\showDOI{\tempurl}


\bibitem[\protect\citeauthoryear{Zhao, Kim, Herman, Pfister, Lau, Echevarria,
  and Bylinskii}{Zhao et~al\mbox{.}}{2020}]%
        {zhao2020IconGen}
\bibfield{author}{\bibinfo{person}{Nanxuan Zhao}, \bibinfo{person}{Nam~Wook
  Kim}, \bibinfo{person}{Laura~Mariah Herman}, \bibinfo{person}{Hanspeter
  Pfister}, \bibinfo{person}{Rynson~W.H. Lau}, \bibinfo{person}{Jose
  Echevarria}, {and} \bibinfo{person}{Zoya Bylinskii}.}
  \bibinfo{year}{2020}\natexlab{}.
\newblock \showarticletitle{ICONATE: Automatic Compound Icon Generation and
  Ideation}. In \bibinfo{booktitle}{\emph{Proceedings of the 2020 CHI
  Conference on Human Factors in Computing Systems}}
  \emph{(\bibinfo{series}{CHI '20})}. \bibinfo{publisher}{Association for
  Computing Machinery}, \bibinfo{address}{New York, NY, USA},
  \bibinfo{pages}{1–13}.
\newblock
\showISBNx{9781450367080}
\urldef\tempurl%
\url{https://doi.org/10.1145/3313831.3376618}
\showDOI{\tempurl}


\bibitem[\protect\citeauthoryear{Zhu, Park, Isola, and Efros}{Zhu
  et~al\mbox{.}}{2017}]%
        {Zhu_2017_ICCV}
\bibfield{author}{\bibinfo{person}{Jun-Yan Zhu}, \bibinfo{person}{Taesung
  Park}, \bibinfo{person}{Phillip Isola}, {and} \bibinfo{person}{Alexei~A.
  Efros}.} \bibinfo{year}{2017}\natexlab{}.
\newblock \showarticletitle{Unpaired Image-To-Image Translation Using
  Cycle-Consistent Adversarial Networks}. In
  \bibinfo{booktitle}{\emph{Proceedings of the IEEE International Conference on
  Computer Vision (ICCV)}}.
\newblock


\end{thebibliography}
